\documentclass[usenatbib,natbib2,natbibmnfix]{mn2e}
\usepackage{mathptm}
\usepackage{times}

\usepackage{epsfig}

\title[]
{The Phoenix Project: the Dark Side of Rich Galaxy Clusters}
\author[Gao et al.]
       {L. Gao$^{1,2}$\thanks{Email:lgao@bao.ac.cn},
         J. F. Navarro$^3$,
         C. S. Frenk$^2$,
         A. Jenkins$^2$,
         V. Springel$^{4,5}$,
         S. D. M. White$^6$
\\
$^1$Partner Group of the Max Planck Institute for Astrophysics,
National Astronomical Observatories, Chinese Academy of Sciences,
Beijing, 100012, China \\
$^2$Institute of Computational Cosmology, Department of Physics, University of Durham,Science Laboratories, South Road, Durham DH1
3LE \\
$^3$Department of Physics and Astronomy, University of Victoria, PO
Box 3055 STN CSC, Victoria, BC, V8W 3P6 Canada \\
$^4$Heidelberger Institut f\"{u}r Theoretische Studien,
Schloss-Wolfsbrunnenweg 35, 69118 Heidelberg, Germany\\
$^5$Zentrum f\"ur Astronomie der Universit\"at Heidelberg,
Astronomisches Recheninstitut, M\"{o}nchhofstr. 12-14, 69120
Heidelberg, Germany\\
$^6$Max-Planck Institute for Astrophysics, Karl-Schwarzschild Str. 1,
D-85748, Garching, Germany \\
}
\begin{document}
\label{firstpage} \maketitle

\begin{abstract}
  We introduce the Phoenix Project, a set of $\Lambda$CDM simulations
  of the dark matter component of nine rich galaxy clusters. Each
  cluster is simulated at least at two different numerical
  resolutions. For eight of them, the highest resolution corresponds
  to $\sim 130$ million particles within the virial radius, while for
  one this number is over one billion. We study the structure
  and substructure of these systems and contrast them with six
  galaxy-sized dark matter haloes from the Aquarius Project, simulated
  at comparable resolution. This comparison highlights the approximate
  mass invariance of CDM halo structure and substructure. We find
  little difference in the spherically-averaged mass,
  pseudo-phase-space density, and velocity anisotropy profiles of
  Aquarius and Phoenix haloes. When scaled to the virial properties of
  the host halo, the abundance and radial distribution of subhaloes are
  also very similar, despite the fact that Aquarius and Phoenix haloes
  differ by roughly three decades in virial mass. The most notable
  difference is that cluster haloes have been assembled more recently
  and are thus significantly less relaxed than galaxy haloes, which
  leads to decreased regularity, increased halo-to-halo scatter and
  sizable deviations from the mean trends. This accentuates the
  effects of the strong asphericity of individual clusters on surface
  density profiles, which may vary by up to a factor of three at a given
  radius, depending on projection. The high apparent concentration
  reported for some strong-lensing clusters might very well reflect
  these effects. A more recent assembly also explains why substructure
  in some Phoenix haloes is slightly more abundant than in Aquarius,
  especially in the inner regions. Resolved subhaloes nevertheless
  contribute only $11 \pm 3\%$ of the virial mass in Phoenix
  clusters. Together, the Phoenix and Aquarius simulation series
  provide a detailed and comprehensive prediction of the cold dark
  matter distribution in galaxies and clusters when the effects of
  baryons can be neglected.
\end{abstract}

\title{The Phoenix project}

\section{Introduction}

The past two decades have witnessed the emergence of a paradigm for
the origin of structure in the Universe. There is now strong evidence
that the dominant forms of the matter-energy content are a combination
of a mysterious form of ``dark energy'' that governs the late
expansion of the Universe, and ``dark matter'' made up of some
kind of non-baryonic, weakly interacting elementary particle left over
from the Big Bang. Although the exact nature of the dark matter
particle is unknown, astrophysical clues to its identity may
be gained by studying its clustering properties on different
scales. Considerable effort has been devoted to this task, and has led
to the crafting of detailed theoretical predictions, especially for
the case of particles with negligible thermal velocity, the
cornerstone of the popular ``cold dark matter'' (CDM) theory. As a
result, we now understand fairly well: (i) the statistics of CDM
clustering on large scales and its dependence on the cosmological
parameters \citep[e.g.,][]{jenkins98,Springel2006}; (ii) the dynamics
of its incorporation into non-linear units (``haloes'') \citep[see,
e.g.,][and references therein]{Wang2011}; and, at least empirically,
(iii) its spatial distribution within such virialized
structures \citep[e.g.,][]{frenk85,NFW96,NFW97,Navarro04,Navarro10}.

Progress in this field has been guided by N-body simulations of ever
increasing numerical resolution and dynamic range
\citep[e.g.][]{frenk85,NFW97,moore99,js02,Navarro04,Gao04a,Diemand04b,Diemand07a,Gao08,sp08b,Stadel09,Navarro10}.
These simulations are essential to investigate highly non-linear
scales such as the haloes of individual galaxies and galaxy groupings,
where simple analytical approximations fail. A few key properties of
CDM haloes are now widely agreed upon, at least when the effects of
baryons are neglected: (a) the presence of a central density ``cusp'';
(b) strong deviations from spherical symmetry; (c) a remarkable
similarity in the shape of the mass profiles; and (d) the presence
of abundant substructure in the form of self-bound ``subhaloes''.

On the scale of individual galaxies, these key predictions have been
confirmed and refined by the latest simulation work, in particular the
Via Lactea simulation series \citep{Diemand07a}, the {\small GHALO}
simulation \citep{Stadel09} and the Aquarius
Project \citep{sp08a,sp08b,Navarro10}. For example, the central
density cusp is now accepted to be shallower than hypothesized in some
earlier work and mass profiles have been shown to be only
approximately self-similar. Further, it is now clear that although
subhaloes are subdominant in terms of total mass, they are still dense
and abundant enough to dominate the dark matter annihilation radiation from a halo.

As shown by \citet{sp08b}, the latter statement requires a
detailed characterization of the substructure, including the internal
properties of the subhaloes, their mass function, and their spatial
distribution within the main halo. The Aquarius Project has provided
compelling, if mainly empirical, guidance to each of these issues in
the case of haloes similar to that of the Milky Way. For example, the subhalo mass
function is well approximated by a power law, $dN/dM\propto
M^{-1.9}$, with a normalisation, in scaled units, weakly dependent on 
halo mass \citep{Gao2010}.
In addition, subhaloes tend to avoid the central region of
the main halo and are more prevalent in the outer
regions. Interestingly, their spatial distribution appears independent
of subhalo mass, a result that, if generally applicable, simplifies
substantially the characterization of substructure. Finally, the
internal structure of subhaloes obeys scaling laws similar to those of
haloes in isolation but slightly modified by the effects of the tidal
field of the main halo: subhaloes are ``denser'', reaching their
peak circular velocity at radii roughly half that of their
isolated counterparts.

Galaxy clusters are a promising venue for testing these
predictions. The central cusp, for example, can be constrained by
combining measurements of the stellar kinematics of the central galaxy
with a lensing analysis of radial and tangential ``arcs'' near the
cluster centre \citep[e.g.,][]{Sand02, Sand04, Meneghetti07, Newman09,
  Zitrin11}. Outside the very centre, the cluster mass profile can be
measured through weak lensing \citep[see,
e.g.,][]{Okabe2010,Oguri2011,Umetsu2011}, X-ray studies of the hot intracluster
medium \citep[ICM; e.g, ][]{Buote2007}, and, more recently, through
the ICM Sunyaev-Ze'ldovich effect on the cosmic microwave background
\citep[see, e.g.,][]{Gralla2011}. In many cases, including
substructure seems {\it required} in order to obtain acceptable fits
\citep[e.g.][]{mao98,mao04,xu09,Natarajan2007,Natarajan2009}, implying
that it should be possible to contrast observations directly with the
CDM substructure predicted by simulations.

Such endeavour has so far been hindered by the lack of
ultra-high-resolution dark matter simulations of galaxy clusters
comparable to the Aquarius series. Indeed, the highest-resolution
galaxy cluster simulations published to date have at most of order a
few million particles within the virial radius
\citep[e.g.][]{Jing2000,sp01,Diemand04b,Reed05a}, roughly one thousand
times fewer than the best resolved Aquarius halo. None of these
cluster simulations are thus able to address conclusively issues such
as the structure of the central cusp or the properties of cluster
substructure.

Although it may be tempting to appeal to the nearly self-similar
nature of CDM haloes to extrapolate the Aquarius results to cluster
scales, it is unclear what systematic uncertainties might be
introduced through such extrapolation. Clusters are rare, dynamically
young objects up to one thousand times more massive than individual
galaxies. They thus trace scales where the CDM power spectrum differs
qualitatively from that of galaxies. Precision work demands that the
near self-similarity of dark haloes be scrutinized directly in order to
provide definitive predictions for the CDM paradigm on cluster
scales. 

To this aim, we have carried out a suite of simulations designed to
address these issues in detail. The Phoenix Project follows
the design of the Aquarius Project and consists of zoomed-in
resimulations of individual galaxy clusters drawn from a
cosmologically representative volume. The simulations follow only the
dark matter component of each cluster, and include the first
simulation of a cluster-sized halo with more than one billion
particles within the virial radius. Like the Aquarius Project on
galaxy scales, the large dynamic range of these simulations allows us
to probe not only the innermost regions of cluster haloes and thus the
structure of the central cusp, but also the statistics, internal
structure, and spatial distribution of cluster substructure over a
mass range spanning seven decades. 

Our paper is organized as follows. In Section~\ref{SecNumExp}, we
describe our numerical techniques and introduce the simulation set. In
Sec.~\ref{SecStruc} and Sec.~\ref{SecSubs} we discuss, respectively,
the density profile and substructure properties of Phoenix haloes
and compare them with those of Aquarius. Sec.~\ref{SecConc} summarizes
our main conclusions. 

\begin{table*}
\begin{tabular}{lccrrccccrr}
\hline
Name & $m_{\rm p}$ & $M_{\rm 200}$ & $r_{\rm 200}$ &$N_{\rm 200}$
&$\epsilon$ & $r_{\rm conv}$\\
   & $[h^{-1}{\rm M}_\odot]$ & $[h^{-1}{\rm M}_\odot]$ & $[h^{-1}{\rm Mpc}]$
   & &$[h^{-1}{\rm kpc}]$&$[h^{-1}{\rm kpc}]$ \\
\hline
Ph-A-1  & $6.355 \times 10^{5}$ & $6.560 \times 10^{14}$  & $1.413$ &$1,032,269,120$ & $0.15$ &$1.2$ &\\
Ph-A-2  & $5.084 \times 10^{6}$ & $6.570 \times 10^{14}$ & $1.414$ &$129,235,472$ & $0.32$ &$2.7$ &\\
Ph-A-3  & $1.716 \times 10^7$ & $6.566 \times 10^{14}$ & $1.413$ &$38,261,560$ & $0.7$ &$4.2$ & \\
Ph-A-4  & $1.373 \times 10^8$ & $6.593 \times 10^{14}$ & $1.415$ &$4,802,516$ & $2.8$ &$9.4$ &\\
\hline
Ph-B-2 &$6.127 \times 10^{6}$ & $8.255 \times 10^{14}$ & $1.526$ &$134,718,112$ & $0.32$ &$3.0$ &\\
Ph-B-4 &$1.656 \times 10^{8}$ & $8.209 \times 10^{14}$ & $1.522$ &$4,956,688$ &$2.8$ &$10.7$ &\\
\hline 
Ph-C-2 &$4.605 \times 10^{6}$ & $5.495 \times 10^{14}$ & $1.386$ &$119,324,008$ &$0.32$ &$2.6$ &\\
Ph-C-4 &$1.182 \times 10^8$ & $5.549 \times 10^{14}$ &$1.383$ &$4,696,046$ &$2.8$ &$9.2$ &\\
\hline
Ph-D-2 &$4.721 \times 10^6$ & $6.191 \times 10^{14}$ & $1.386$ &$130,529,200$ &$0.32$ &$2.7$ &\\
Ph-D-4 &$1.373 \times 10^8$ & $6.162 \times 10^{14}$ & $1.384$ &$4,488,330$ &$2.8$ &$9.4$ &\\
\hline
Ph-E-2 &$4.425 \times 10^6$ & $5.969 \times 10^{14}$ & $1.369$ &$130,529,200$ &$0.32$ &$2.4$ &\\
Ph-E-4 &$1.017 \times 10^8$ & $5.923 \times 10^{14}$ & $1.366$ &$5,824,375$ &$2.8$ &$8.4$ &\\
\hline
Ph-F-2 &$4.425 \times 10^6$ & $7.997 \times 10^{14}$ & $1.509$ &$129,221,216$ &$0.32$ &$2.8$ &\\
Ph-F-4 &$1.682 \times 10^8$ & $8.039 \times 10^{14}$ & $1.512$ &$4,779,008$ &$2.8$ &$10.3$ &\\
\hline
Ph-G-2 &$8.599 \times 10^6$ & $1.150 \times 10^{15}$ & $1.704$ &$133,730,958$ &$0.32$ &$3.2$ &\\
Ph-G-4 &$2.907 \times 10^8$ & $1.148 \times 10^{15}$ & $1.703$ &$3,949,310$ &$2.8$ &$13.1$ &\\
\hline
Ph-H-2 &$8.600 \times 10^6$ & $1.136 \times 10^{15}$ & $1.686$ &$129,488,456$ &$0.32$ &$2.9$ &\\
Ph-H-4 &$2.502 \times 10^8$ & $1.150 \times 10^{15}$ & $1.686$ &$4,456,720$ &$2.8$ &$11.8$ &\\
\hline
Ph-I-2 &$1.841 \times 10^7$ & $2.411 \times 10^{15}$ & $2.185$ &$131,845,620$ &$0.32$ &$2.9$ &\\
Ph-I-4 &$4.559 \times 10^8$ & $2.427 \times 10^{15}$ & $2.181$ &$5,289,259$ &$2.8$ &$14.2$ &\\
\hline
\end{tabular}
\caption{Basic parameters of the Phoenix simulations. Each of the nine haloes is
  labelled as Ph-X-N, where the letter X (from A to I) identifies each
  halo, and N, which runs from 1 to 4, refers to the numerical resolution (1
  is highest). The parameter  $m_{\rm p}$
  gives the particle mass in the high-resolution region that includes
  the cluster; $M_{\rm 200}$ is the virial mass of
  the halo; $r_{\rm 200}$ is the
  corresponding virial radius; and $N_{200}$ states the number of
  particles inside $r_{200}$. The parameter $\epsilon$ is the  Plummer-equivalent
  gravitational softening length, so that pairwise interactions are
  fully Newtonian when separated by a distance greater than $2.8\,
  \epsilon$. The last column lists the ``convergence radius'', $r_{\rm
    conv}$, outside which the circular velocity is expected to
  converge to better than $10\%$.
  \label{TabSimParam}}
\end{table*}

\begin{figure*}
\hspace{0.13cm}\resizebox{16cm}{!}{\includegraphics{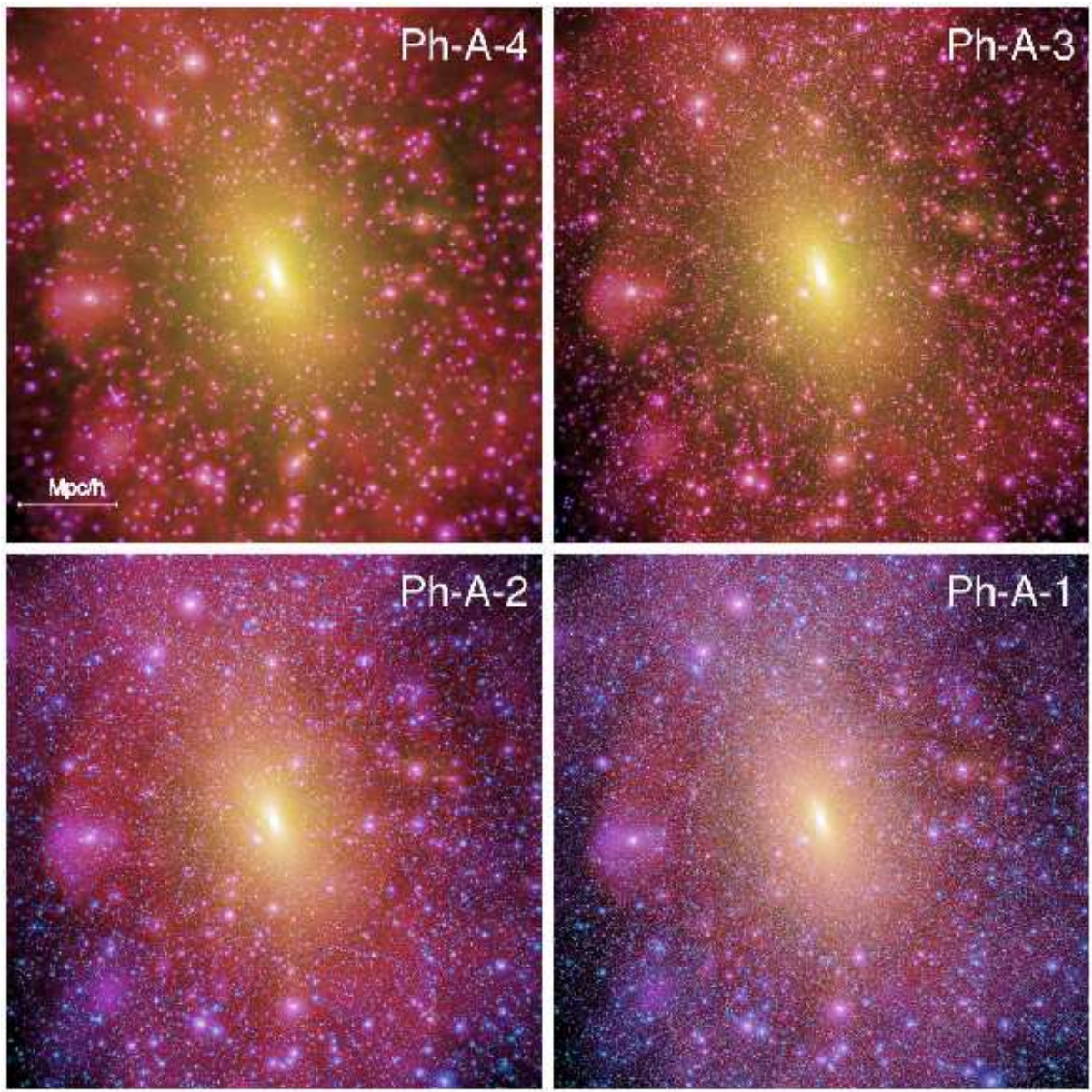}}
\caption{Images of cluster Ph-A at four different numerical
  resolutions. Each panel projects a cubic volume $5 \, h^{-1}$ Mpc on a
  side. The brightness of each image pixel is proportional to the
  logarithm of the square of the dark matter density projected along
  the line of sight, and the hue encodes the local velocity dispersion
  density-weighted along the line of sight (see text for
  details). This rendering choice highlights the presence of
  substructure which, although abundant, contributes less than about
  $10\%$ of the total mass within the virial
  radius.} 
\label{FigPhADiffRes}
\end{figure*}

\begin{figure*}
\hspace{0.13cm}\resizebox{16cm}{!}{\includegraphics{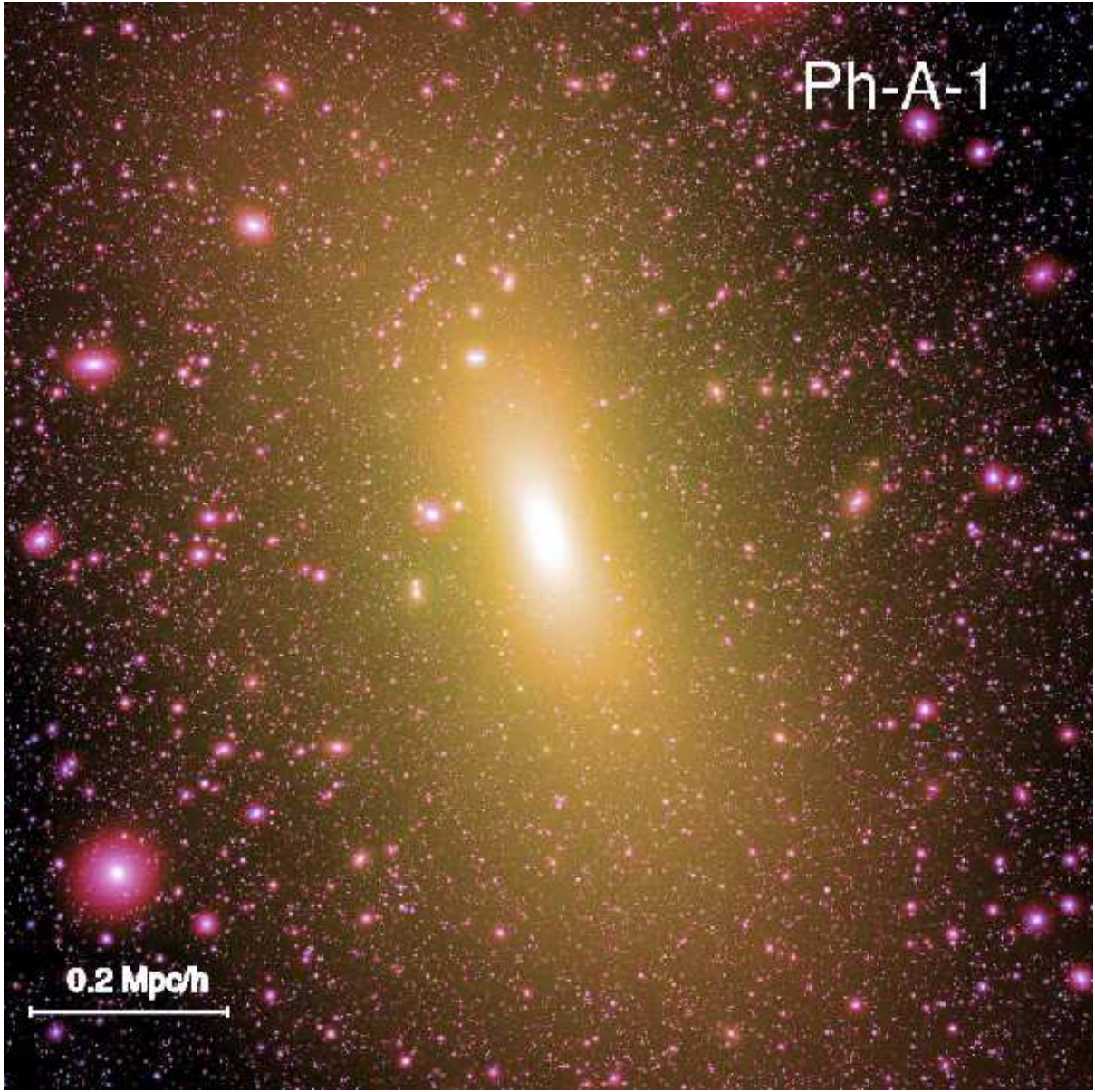}}
\caption{The inner $\sim 1 \, h^{-1}$ Mpc of Ph-A-1. Color coding is as in
  Fig.~\ref{FigPhADiffRes}. This figure illustrates clearly the strong
  asphericity of the halo; the presence of several nested levels of
  substructure, and the tendency of subhaloes to avoid the halo
  centre.} 
\label{FigPhA1}
\end{figure*}

\begin{figure*}
\hspace{0.13cm}\resizebox{16cm}{!}{\includegraphics{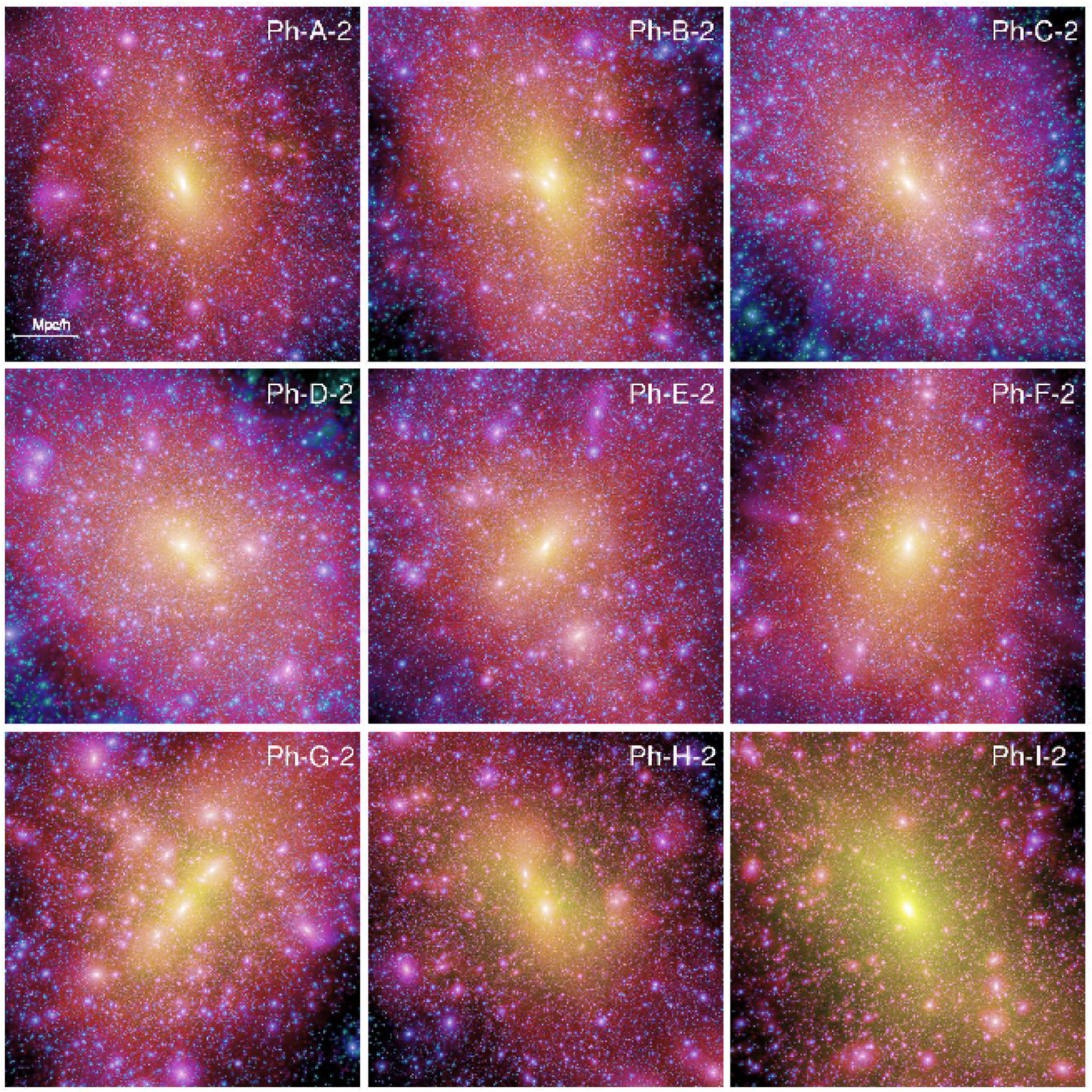}}
\caption{As Fig.~\ref{FigPhADiffRes}, but for all level-2 Phoenix
  clusters at $z=0$. Boxes are all $5 \, h^{-1}$ Mpc on a side. Note that
  the appearance of several Phoenix clusters is suggestive of a
  transient evolutionary stage, characterized by the presence of a
  number of undissolved substructure groupings. Ph-G-2 is a particularly good
  example of this irregular structure which may be traced to its
  recent assembly time; this cluster has acquired half its mass since
  $z=0.18$.} \label{FigImAllClus}
\end{figure*}
\section{The simulations}
\label{SecNumExp}

The Phoenix Project consists of a series of simulations of $9$
different galaxy clusters with masses exceeding $5\times 10^{14} \,
h^{-1} \, M_\odot$. These clusters were selected from a large
cosmological volume and resimulated individually at varying
resolution. Details of the resimulation procedure are given below.

\subsection{Cosmology}
\label{SecCosm}

All the simulations reported here adopt the cosmological parameters of
the Millennium Simulation \citep{sp05}: $\Omega_{\rm M}=0.25$,
$\Omega_{\Lambda}=0.75$, $\sigma_8=0.9$, $n_s=1$, and a present-day
value of the Hubble constant $H_0=100\,h\,$km $s^{-1}$ Mpc$^{-1} =73 $
km s$^{-1}$ Mpc$^{-1}$. This is also the set of cosmological
parameters adopted for the Aquarius project \citep{sp08a}, which
targeted haloes a thousand times less massive.  Although they are
inconsistent with the latest {\small CMB} data \citep{Komatsu2011} the
differences are not large (the main difference is that a lower value
of $\sigma_8=0.81$ is now preferred) and they are expected to affect
only the abundance of cluster haloes rather than their detailed
structure and substructure properties \citep{Wang2012}. This choice also has the
advantage that any difference between Aquarius and Phoenix haloes can
be traced to the different mass scales and not to variations in the
cosmological model.

\subsection{Cluster Sample and Resimulations}
\label{SecParSim}

The Phoenix cluster sample is selected for resimulation from the
Millennium Simulation friends-of-friends group catalog at $z=0$. Six
clusters were selected at random from the $72$ systems with
virial{\footnote{We define the virial radius of a cluster, $r_{200}$,
    as that of a sphere of mean density $200$ times the critical
    density for closure; $\rho_{\rm crit}=3H_0^2/8\pi G$. The virial
    radius defines implicitly the virial mass of a cluster, $M_{200}$,
    and its virial velocity, $V_{200}=\sqrt{GM_{200}/r_{200}}$.}}
mass in the range $5<M_{200}/10^{14}\, h^{-1} \,M_{\odot}<10$. In
order to sample the tail of rare rich clusters three further Phoenix
clusters were selected from the nine Millennium halos which have
$M_{200} > 10^{15} h^{-1} M_{\odot}$.

The initial conditions for resimulation were set up using a procedure
analogous to that used for the Aquarius haloes and described in detail
by \citet{Power2003} and \citet{sp08b}. The only difference is
that the initial displacements and velocities were computed using
second-order Lagrangian perturbation theory, as described by
\citet{Jenkins10}. All nine haloes were resimulated at least twice
using different numerical resolution (level 2 and level 4,
respectively). At level 2 each cluster has between $120$ and $135$
million particles within the virial radius; at level 4 each system is
made up of 4 to 6 million particles.

We have selected one of the clusters (Ph-A) for a numerical resolution
study and have carried out an extra level-3 run (with roughly $40$
million particles within $r_{200}$) and a flagship level-1 run, where
we followed $4.05$ billion high-resolution particles in total,
$1.03$ billion of which are found within $r_{200}$ at $z=0$. For ease
of reference we label the runs using the convention Ph-X-N, where X is a
letter from A to I that identifies each individual cluster
and N is a number from 1 to 4 that specifies the resolution level. The
simulation parameters are summarized in Table~\ref{TabSimParam}.
We have used for all runs the P-Gadget-3 code, a version of Gadget-2
\citep{sp05} especially optimized for zoomed-in cosmological
resimulations in distributed-memory massively-parallel computers. The
code is identical to that used for the Aquarius Project \citep{sp08a}.
The simulations were carried out on Deepcomp 7000 at the
Supercomputer Center of the Chinese Academy of Science. The largest
simulation, Ph-A-1, used 3 Tbs of memory on 1024 cores  and took
about $1.9$ million CPU hours. The initial conditions were
generated at the Institute for Computational Cosmology (Durham
University).

The gravitational softening of each run was chosen following the
``optimal'' prescription of \citet{Power2003}. It is kept fixed in
comoving coordinates throughout each run and is listed in Table
~\ref{TabSimParam}. Our highest-resolution run (Ph-A-1) has a nominal
(Plummer-equivalent) spatial resolution of just $150\, h^{-1}$ pc.

Haloes are identified in each run using the friends-of-friends ({\small
  FOF}) group finding algorithm with linking length set to $20\%$ of
the mean interparticle separation \citep{Davis85}. Substructure
within {\small FOF} haloes is identified by {\small SUBFIND}
\citep{Springel01a}, a groupfinding algorithm that searches
recursively for self-bound subhaloes. Both {\small FOF} and {\small
  SUBFIND} have been integrated within P-Gadget-3 and are run
on-the-fly each time a simulation snapshot is created.

We have stored for each run $72$ snapshots uniformly spaced in
$\log_{10} a$, starting at $a=0.017$ ($a=1/(1+z)$ is the expansion
factor). The initial conditions are set at $z_{\rm init}= 63$ {\bf for
  our level $4$ and at $z_{\rm init}=79$ for the rest}. The
large number of outputs is designed to allow us in future work to
implement semi-analytic models of galaxy formation in order to follow
the evolution of the baryonic component of galaxies within rich clusters.

We list the basic structural parameters of Phoenix clusters at
redshift $z=0$ in Table~\ref{TabHaloParam}. These include the peak
circular velocity, $V_{\rm max}$, and the radius, $r_{\rm max}$, at
which it is reached; the half-mass formation redshift, $z_{\rm h}$,
when the main progenitor first reaches half the final halo mass; the
concentration parameters, $c$ and $c_{\rm E}$, obtained from the
best-fit NFW \citep{NFW96,NFW97} and Einasto (1965) profiles,
respectively; the figure of merit, $Q_{\rm NFW}$ and $Q_{\rm E}$,
associated with each of those fits; and the Einasto ``shape''
parameter $\alpha$. (See the Appendix for definitions corresponding to
these fitting formulae and for details on the profile-fitting
procedure.) $N_{\rm sub}$ is the total number of subhaloes with more
than $20$ particles identified by {\small SUBFIND} inside $r_{200}$;
$f_{\rm sub}$ is the total mass contributed by these subhaloes,
expressed as a fraction of the virial mass. 

\section{The Structure of Phoenix Clusters}
\label{SecStruc}

\begin{figure*}
\hspace{0.13cm}\resizebox{8cm}{!}{\includegraphics{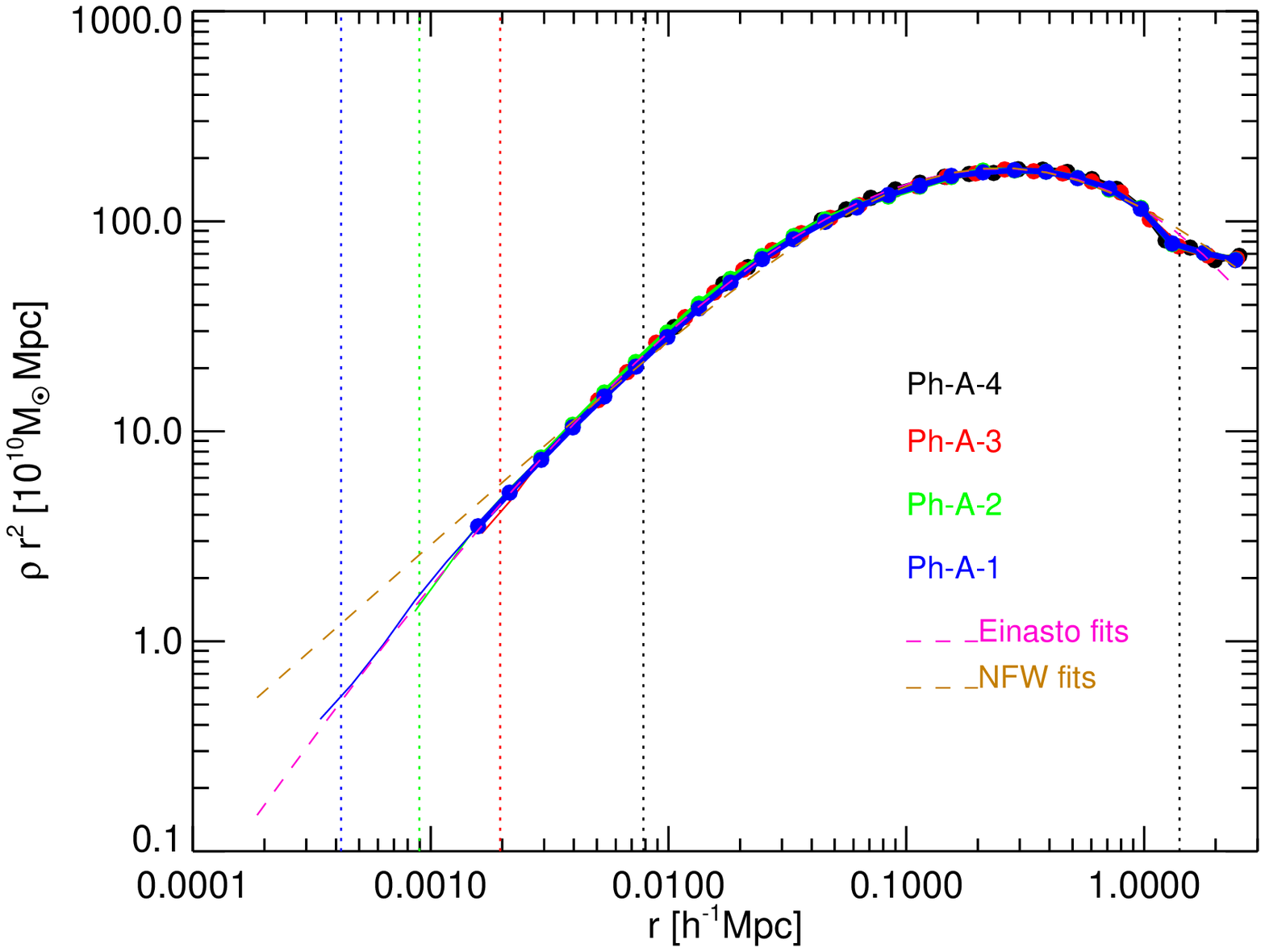}}
\hspace{0.13cm}\resizebox{8cm}{!}{\includegraphics{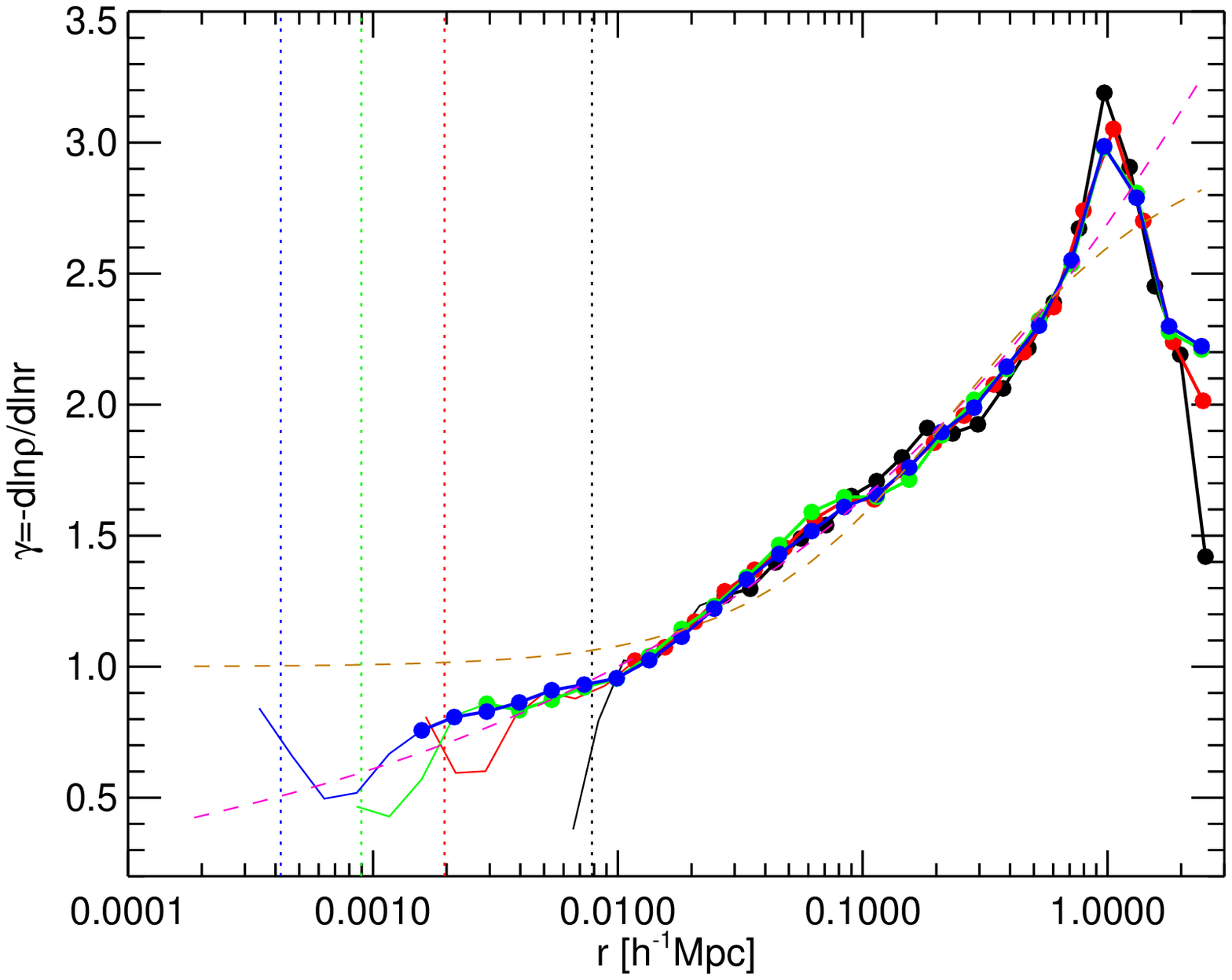}}
\caption{{\it Left panel}: Spherically-averaged density profile of
  halo Ph-A at $z=0$. Different colors correspond to the four different
  resolution runs listed in Table~\ref{TabSimParam}. The panel on the
  left shows the density multiplied by $r^2$ in order to enhance the
  dynamic range of the plot. Each profile is shown with a thick line
  connecting filled circles from the ``convergence radius'', $r_{\rm
    conv}$, outwards \citep{Power2003}.  Thin curves extend the
  profiles inwards down to $r=2\, \epsilon$, where $\epsilon$ is the
  Plummer-equivalent gravitational softening length. Vertical dotted
  lines indicate, for each run, $2.8\, \epsilon$, the distance beyond
  which pairwise particle interactions are fully Newtonian.  Note the
  excellent numerical convergence achieved for each simulation outside
  their $r_{\rm conv}$. An NFW profile with concentration
  $c=5.63$ (thin dashed brown line) and an Einasto profile with
  $\alpha=0.22$ and $c_{\rm E}=5.59$ (thin dashed magenta line) are also
  shown for comparison. {\it Right panel}: Logarithmic slope
  ($\gamma=-d\ln \rho/d\ln r$) of the density profile as a function of
  radius. Colors and line types are the same as in the left
  panel. Note again the excellent convergence achieved in all runs
  at radii outside the convergence radius, $r_{\rm conv}$.}
  \label{FigConvDensProf}
\end{figure*}

We shall focus our analysis on the properties of Phoenix clusters at
$z=0$. Figure~\ref{FigPhADiffRes} shows Ph-A at the four different
numerical resolutions. As in \citet{sp08b}, this and other
cluster images are constructed so that the brightness of each pixel is
proportional to the logarithm of the square of the dark matter density
projected along the line of sight, 
\begin{equation} S(x,y) =
  \int\rho_{\rm loc}^2({\bf r}) \, dz
\end{equation}
while the color hue encodes the mean dark matter velocity dispersion, 
\begin{equation}
\sigma(x,y)=\frac{1}{S(x,y)}\int\sigma_{\rm loc}({\bf r}) \, \rho_{\rm  loc}^2({\bf
  r})\, dz
\end{equation}
Here the local dark matter density, $\rho_{\rm loc}(r)$, and the local velocity
dispersion, $\sigma_{\rm loc}(r)$, are estimated using an {\small SPH}
kernel interpolation scheme. 

Figure~\ref{FigPhADiffRes} shows that the main result of increasing
the number of particles is the ability to resolve larger numbers of
subhaloes. On the other hand, the main properties of the cluster, such as its
shape and orientation, the overall mass profile, and even the location
of the largest subclumps remain invariant in all four Ph-A
realizations.

Fig.~\ref{FigPhA1} is analogous to Fig.~\ref{FigPhADiffRes}, but for
the inner $\sim 1 \, h^{-1}$ Mpc of Ph-A-1 (our highest resolution
run). This image highlights the strong asphericity of the halo, as
well as the presence of several nested levels of substructure (i.e.,
subhaloes within subhaloes). It also shows that subhaloes tend to avoid
the central regions. These characteristics are shared with galaxy-sized
haloes \citep{sp08b}, and appear to be typical of CDM haloes on
all mass scales.

Fig.~\ref{FigImAllClus} is analogous to Fig.~\ref{FigPhADiffRes} but
for all level-2 Phoenix haloes at $z=0$. This figure shows that the
main characteristics of Ph-A described above are common to all Phoenix
clusters: strong asphericity; abundant substructure; and a marked
difference between the spatial distribution of mass (which is highly
concentrated) and that of subhaloes (which tend to avoid the central
regions). 

Fig.~\ref{FigImAllClus} also highlights an important characteristic of
cluster-sized dark matter haloes: the presence of ``multiple centres''
traced by groups of subhaloes, as well as the overall impression that
many systems are in a transient, unrelaxed stage of their
evolution. This is expected, given the late assembly of the clusters:
Ph-G-2, for example, assembled half its final mass after $z=0.18$; the
median half-mass assembly redshift for all Phoenix clusters is just
$z=0.56$. Ph-A, on the other hand, appears relaxed; this cluster
has the highest formation redshift of our sample, $z_{\rm h}\sim
1.2$. 

\begin{table*}
\begin{tabular}{lccrrrccccrrr}
\hline
Name & $V_{\rm max}$ & $r_{\rm max}$ &$z_{\rm h}$ &$c_{\rm
  E}$ &$c$ &$Q_{\rm E}$ &$Q_{\rm NFW}$ &$\alpha$
&$N_{\rm sub}$ &$f_{\rm sub}$ &$d_{\rm off}$ \\

   & [km s$^{-1}$] & $[h^{-1}{\rm Mpc}]$ &&&\\
\hline
Ph-A-1  & $1521.82$  &$0.55$  &$1.17$ &$5.59$ &$5.63$ &$0.037$ &$0.093$ &$0.215$ &$192,206$ &$0.080$ &$0.04$ \\
Ph-A-2  & $1527.24$ &$0.55$ &$1.17$ &$5.72$ &$5.96$ &$0.039$ &$0.075$ &$0.216$ &$26,896$ &$0.071$ &$0.04$\\
Ph-A-3  & $1529.41$ &$0.56$ &$1.17$ &$5.69$ &$6.04$ &$0.038$ &$0.061$ &$0.218$ &$8,478$ &$0.062$ &$0.04$\\
Ph-A-4  & $1538.88$ &$0.59$ &$1.17$ &$5.71$ &$6.14$ &$0.052$ &$0.063$ &$0.219$ &$1,049$ &$0.049$ &$0.04$\\
\hline
Ph-B-2 &$1624.52$ &$0.53$ &$0.46$ & $4.41$ &$4.19$ &$0.127$ &$0.108$ &$0.235$& $38,659$ &$0.108$ &$0.02$\\
Ph-B-4 &$1623.12$ &$0.56$ &$0.46$ & $4.40$ &$4.06$ &$0.107$ &$0.117$ &$0.276$ &$1,657$  &$0.081$ &$0.02$\\
\hline
Ph-C-2 &$1294.19$ &$0.65$ &$0.76$ &$4.27$ &$5.11$  &$0.077$ &$0.104$ &$0.181$ &$33,529$ &$0.114$ &$0.06$\\
Ph-C-4 &$1310.19$ &$0.78$ &$0.76$ &$4.34$ &$4.72$  &$0.085$ &$0.112$ &$0.185$ &$1,489$  &$0.095$ &$0.06$\\
\hline
Ph-D-2 &$1393.13$ &$0.68$ &$0.46$ &$3.88$ &$4.08$  &$0.122$ &$0.086$ &$0.205$ &$38,199$ &$0.124$ &$0.05$\\
Ph-D-4 &$1436.10$ &$0.65$ &$0.46$ &$4.03$ &$4.34$  &$0.136$ &$0.127$ &$0.212$ &$1,491$  &$0.093$ &$0.05$\\
\hline
Ph-E-2 &$1385.78$ &$0.65$ &$0.91$ &$3.48$ &$5.19$  &$0.067$ &$0.135$ &$0.149$ &$33,678$ &$0.101$ &$0.04$\\
Ph-E-4 &$1399.96$ &$0.68$ &$0.91$ &$4.02$ &$4.82$  &$0.048$ &$0.079$ &$0.181$ &$1,547$ &$0.070$ &$0.04$\\
\hline
Ph-F-2 &$1543.27$ &$0.60$ &$1.1$ &$3.81$ &$4.61$  &$0.053$ &$0.048$ &$0.186$ &$31,247$ &$0.095$ &$0.05$\\
Ph-F-4 &$1559.44$ &$0.62$ &$1.1$ &$4.00$ &$4.54$  &$0.059$ &$0.057$ &$0.203$ &$1,547$  &$.075$ &$0.05$\\
\hline
Ph-G-2 &$1561.75$ &$1.06$ &$0.18$ &$0.78$ &$3.33$ &$0.100$ &$0.221$ &$0.097$ &$42,528$ &$0.168$ &$0.17$\\
Ph-G-4 &$1599.17$ &$1.04$ &$0.18$ &$1.10$ &$2.98$ &$0.109$ &$0.164$ &$0.116$ &$1,586$  &$0.140$ &$0.17$\\
\hline
Ph-H-2 &$1676.43$ &$1.14$ &$0.21$ &$1.98$ &$4.66$ &$0.155$ &$0.212$ &$0.117$ &$35,048$ &$0.095$ &$0.1$\\
Ph-H-4 &$1710.19$ &$1.14$ &$0.21$ &$2.75$ &$3.59$ &$0.109$ &$0.115$ &$0.178$ &$1,437$  &$0.069$ &$0.1$\\
\hline
Ph-I-2 &$2236.05$ &$1.03$ &$0.56$ &$4.18$ &$4.86$ &$0.041$ &$0.059$ &$0.190$ &$35,754$ &$0.102$ &$0.02$\\
Ph-I-4 &$2269.09$ &$1.05$ &$0.56$ &$4.48$ &$5.02$ &$0.045$ &$0.051$ &$0.208$ &$1,641$  &$0.073$ &$0.02$\\

\hline
\end{tabular}
\caption{Basic structural parameters of Phoenix clusters at $z=0$. The
  leftmost column labels each run, as in Table~\ref{TabSimParam}; the
  second and third columns list the peak circular velocity, $V_{\rm  max}$, and the radius, $r_{\rm max}$, at which it is reached. The concentration parameters of the best NFW \citep{NFW96,NFW97} and Einasto \citep{Einasto65} fits are
  listed under $c$ and $c_{\rm E}$,  respectively. $Q_{\rm
    NFW}$ and $Q_{\rm E}$ are the figures of  merit of the best NFW
  and Einasto fits, respectively. The column labelled  $\alpha$ lists the Einasto shape parameter. $N_{\rm sub}$ denotes the
  total number of subhaloes with more than $20$ particles identified
  within $r_{200}$;  $f_{\rm sub}$ is the fraction of the virial mass contributed by such
  subhaloes; and $d_{\rm off}$ is the distance from the gravitational
  potential minimum to the centre of mass of particles within the
  virial radius, in units of $r_{200}$.
  \label{TabHaloParam}}
\end{table*}

The late assembly and concomitant departures from equilibrium are
characteristics that set clusters apart from galaxy-sized haloes; for
comparison, the median half-mass formation redshift of Aquarius haloes
is $z\sim 2$. Table~\ref{TabHaloParam} lists two quantitative measures
of departures from equilibrium: the fraction of mass in substructures,
$f_{\rm sub}$, and the offset, $d_{\rm off}$, between the centre of
mass of the halo and the location of the potential minimum expressed
in units of the virial radius \citep[for further discussion of these
parameters see][]{Neto07}. These correlate well with the formation
redshift, $z_{\rm h}$, and are significantly larger, on average, than
in the galaxy-sized Aquarius haloes (see Table~\ref{TabCompAqPh}).

\begin{table*}
\begin{tabular}{lccccccccccc}
\hline
Name  &$\langle z_{\rm  h} \rangle$ &$\langle d_{\rm off} \rangle$ &$\langle f_{\rm sub} \rangle$ &$\langle \alpha  \rangle$ & $\langle Q_{\rm  min} \rangle$ &$\langle C_\chi \rangle$&$\langle \chi  \rangle$ &$\langle N_{m} \rangle$ &$\langle s  \rangle$ &$\langle N_v  \rangle$ &$\langle d  \rangle$ \\
\hline
Phoenix  &$0.65$ & $0.061$ &$0.109$ &$0.175$ &$0.086$  &$1.75$
&$-1.86$ &$7866$&$  -0.97$ &$3984$ &$-3.32$ \\
 &$\pm 0.36$& $\pm 0.047$ &$\pm 0.027$ &$\pm 0.046$ &$\pm 0.041$ &$\pm
 0.29$ &$\pm 0.04$ &$\pm 965$  &$\pm 0.02$ &$\pm 317$&$\pm 0.10$ \\
\hline
Aquarius &$1.65$ & $0.032$ &$0.071$ &$0.159$ &$0.048$ &$2.19$ &$-1.82$ &$5092$ &$-0.94$ &$4033$&$-3.13$ \\
 &$\pm 0.65$ & $\pm 0.011$ &$\pm 0.022$ &$\pm 0.022$ &$\pm 0.012$ &$\pm 0.14$ &$\pm 0.02$ &$\pm 677$ &$\pm 0.02$ &$\pm 500$&$\pm 0.09$ \\
\hline
\end{tabular}
\caption{Comparison of the average properties of the six galaxy-sized
  Aquarius haloes and the nine cluster-sized Phoenix haloes. Sample
  averages are listed for each quantity together with the rms dispersion around the
  mean. The first column identifies the simulation set; $z_h$ is the
  half-mass formation redshift; $d_{\rm off}$ and $f_{\rm sub}$ are
  the dynamical relaxation diagnostics introduced in
  Table~\ref{TabHaloParam}; $\alpha$ is the best-fit  Einasto shape parameter and
  $Q_{\rm min}$ the goodness of fit measure (Sec.~\ref{SecApp});
  $C_{\chi}$ and $\chi$ are the parameters of power-law fits to the
  pseudo-phase-space density profile (eq.~\ref{EqPPSD}); $N_m$ and $s$ describe the power-law fits to the subhalo mass function, $N(>\mu)=N_{m} \, (\mu/10^{-6})^{s}$
  (eq.~\ref{EqSubMF}); $N_v$ and $d$ those corresponding to fits of
  the form, $N(>\nu)=N_v\, (\nu/0.025)^d$,  to the subhalo velocity
  function  (eq.~\ref{EqSubVF}).}
  \label{TabCompAqPh}
\end{table*}

\subsection{Mass Profiles}
\label{SecMassProf}

\begin{figure*}
\hspace{0.13cm}\resizebox{8cm}{!}{\includegraphics{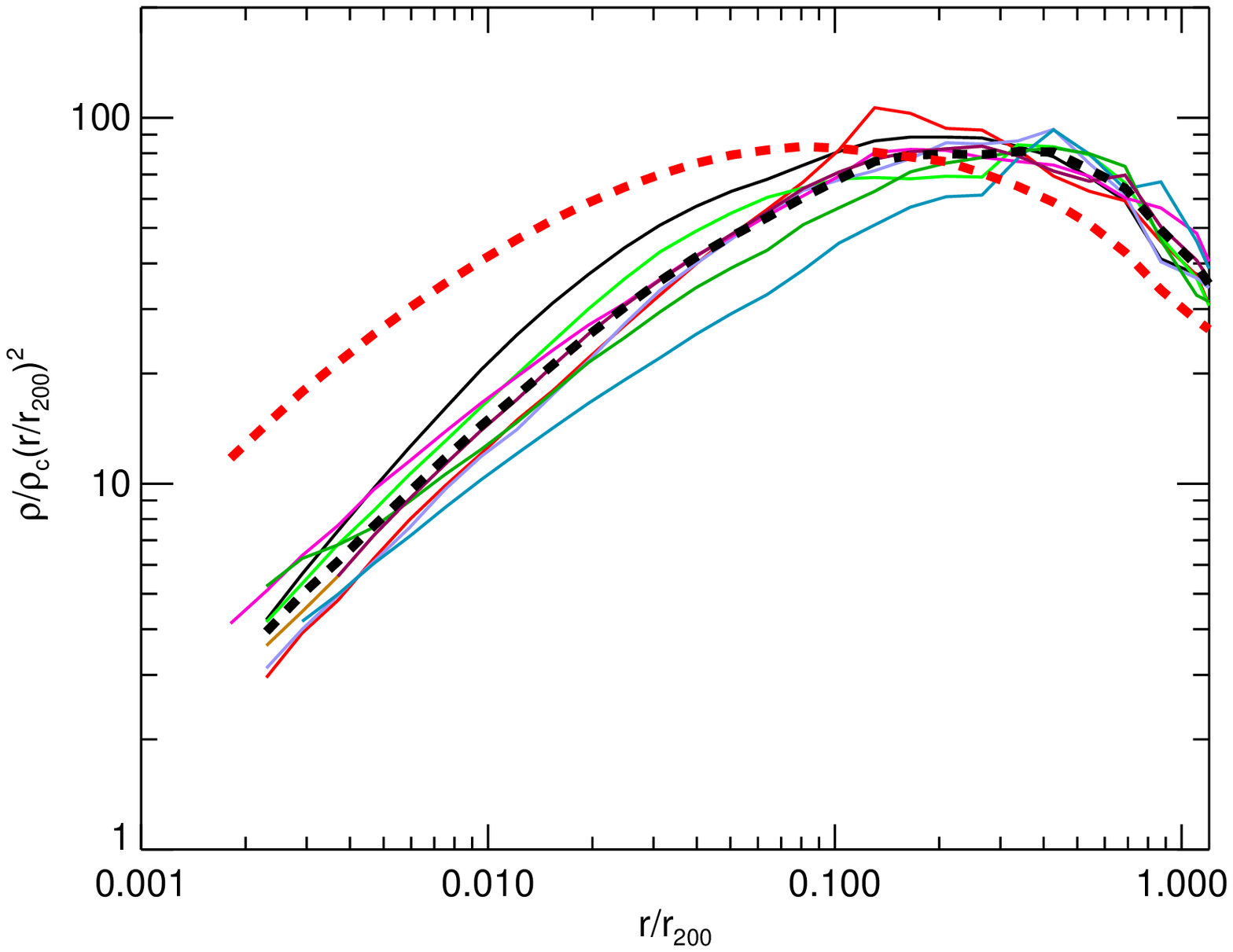}}
\hspace{0.13cm}\resizebox{8cm}{!}{\includegraphics{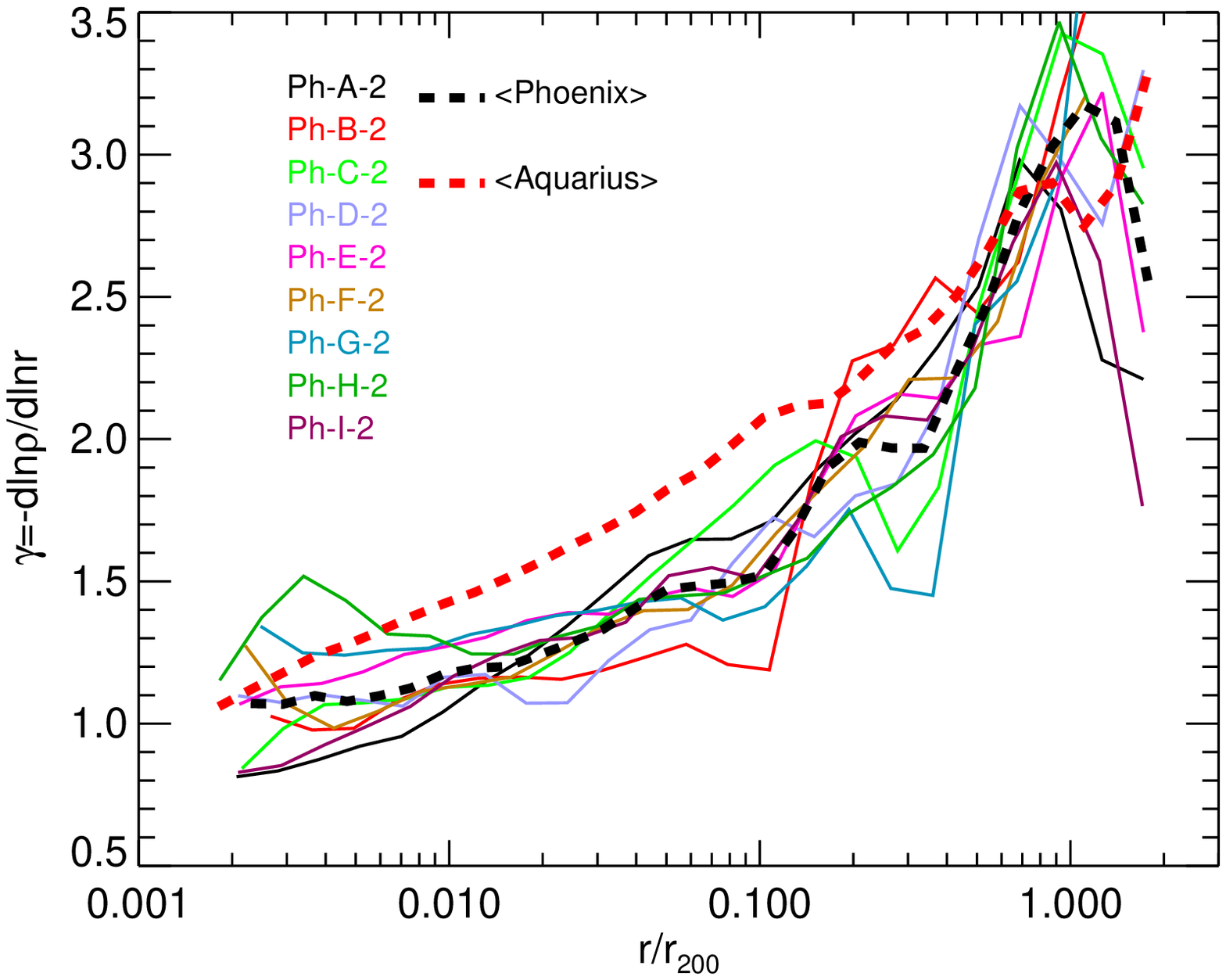}}
\hspace{0.13cm}\resizebox{8cm}{!}{\includegraphics{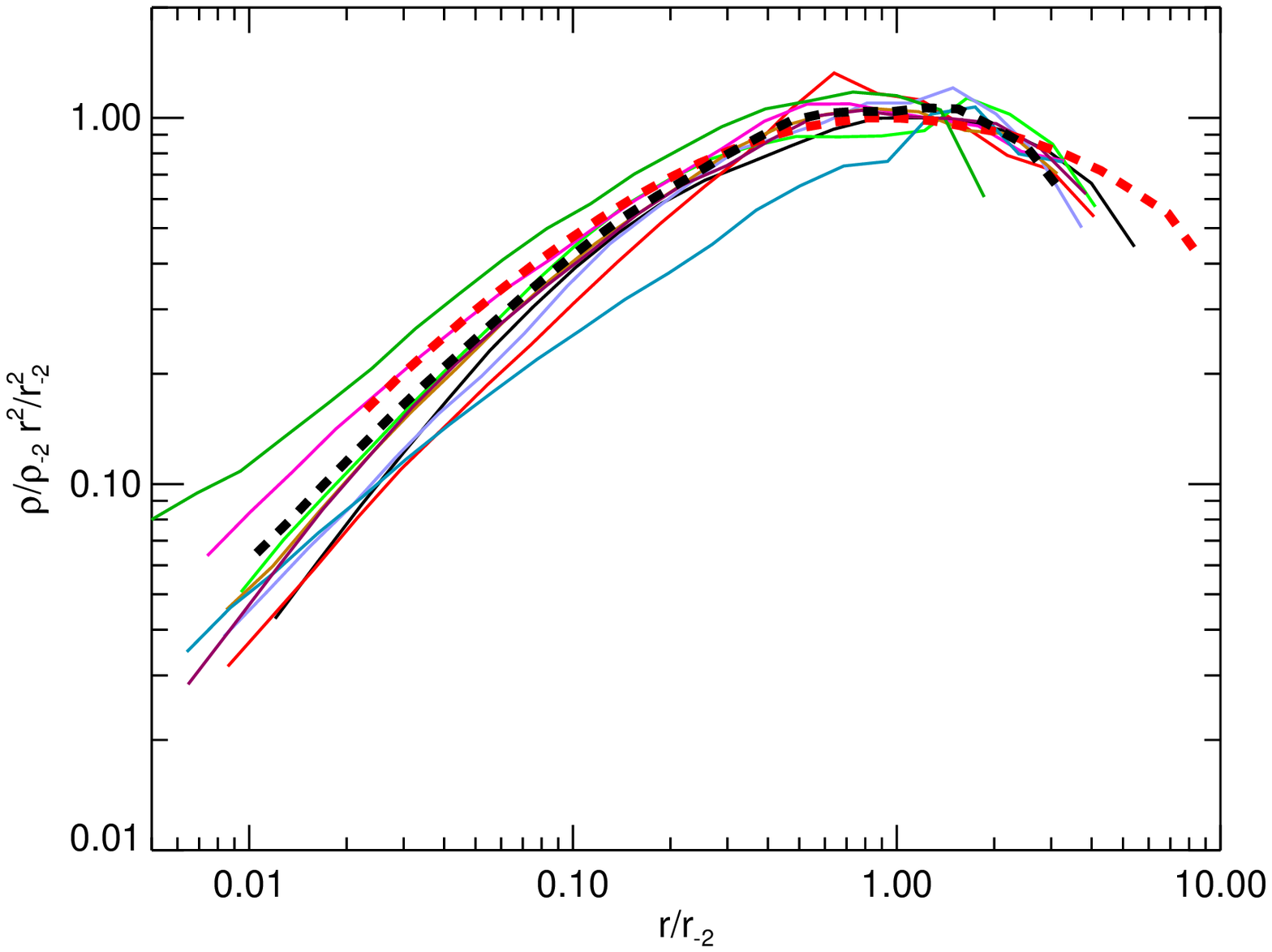}}
\hspace{0.13cm}\resizebox{8cm}{!}{\includegraphics{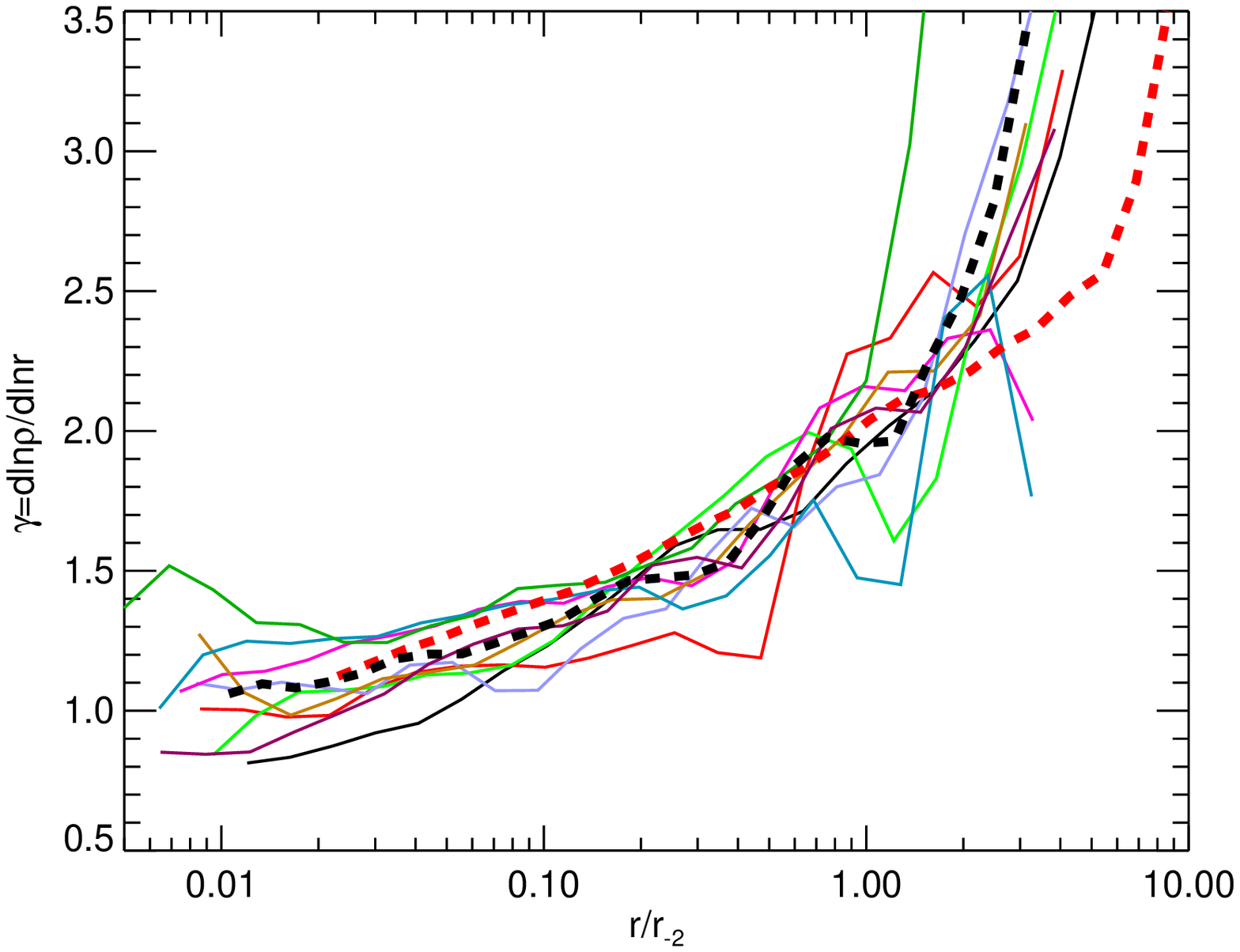}}
\caption{Spherically-averaged density (left panels) and logarithmic
  slope (right panels) of all level-2 Phoenix haloes as a function of
  radius. Radii have been scaled to the virial radius of each halo in
  the top panels and to the ``scale radius'', $r_{-2}$, of the
  best-fit Einasto profile in the bottom panels. Profiles are plotted
  down to the convergence radius, $r_{\rm conv}$.  The thick dashed
  black line shows the average density profile of all Phoenix haloes,
  computed after stacking the nine haloes, each scaled to its own
  virial mass and radius. The thick red dashed line shows the result
  of the same stacking procedure, but applied to the Aquarius haloes.}
\label{FigAllDensProf}
\end{figure*}

We explore in this section the spherically-averaged mass profiles of
Phoenix haloes. We begin by using the four Ph-A realizations to assess
the limitations introduced by finite numerical resolution.
Fig.~\ref{FigConvDensProf} shows the density profile, $\rho(r)$, as
well as the radial dependence of the logarithmic slope, $\gamma=-d\ln
\rho/d \ln r$, for Ph-A-1 through Ph-A-4. As discussed by
\citet{Power2003} and \citet{Navarro10}, the mass profiles of
simulated haloes are robustly determined in regions where the two
body-relaxation time exceeds the age of the Universe. This constraint
defines a ``convergence radius'', $r_{\rm conv}$, outside which the
circular velocity, $V_c=(GM(<r)/r)^{1/2}$, is expected to converge to
better than $10\%$. Since $V_c$ is a cumulative measure we expect
$r_{\rm conv}$ to be a {\it conservative} indicator of the innermost
radius where local estimates of the density, $\rho(r)$, converge to
better than $10\%$.

This is indeed the case for Ph-A, as shown in
Fig.~\ref{FigConvDensProf}. The left panel shows $\rho(r)$, multiplied
by $r^2$ in order to remove the dominant radial trend so as to enhance
the dynamic range of the plot.  Thick lines highlight the radial range
of the profile outside the convergence radius; the density clearly
converges to better than $10\%$ at radii greater than $r_{\rm
  conv}$. In those regions the logarithmic slope $\gamma$ is also
robustly and accurately determined. We conclude that $r>r_{\rm conv}$
is a simple and useful prescription that identifies the regions
unaffected by numerical limitations. We list $r_{\rm conv}$ for all
Phoenix runs in Table~\ref{TabSimParam}.

The thin dashed lines in Fig.\ref{FigConvDensProf} indicate the
best-fit NFW (brown) and Einasto (magenta) profiles, computed as
described in the Appendix. The NFW shape is fixed in this log-log
plot, whereas the Einasto shape is controlled by the parameter
$\alpha$, which is found to be $0.215$ by the fitting procedure when
applied to the Ph-A-1 profile. This figure suggests that the shape of
the mass profile deviates slightly but systematically from the NFW
profile. Although it is possible to obtain excellent fits over the
resolved radial range with the NFW formula (typical residuals are less
than $\sim 10\%$), there is a clear indication that the density
profile near the centre is shallower than the asymptotic $r^{-1}$ NFW
cusp. In agreement with results from the Aquarius Project
\citep{Navarro10}, there is little indication that the central density
cusp of Ph-A is approaching a power-law; the profile becomes gradually
shallower all the way in to the innermost resolved radius. This radial
trend is very well described by the Einasto profile.

\begin{figure*}
\hspace{0.13cm}\resizebox{8cm}{!}{\includegraphics{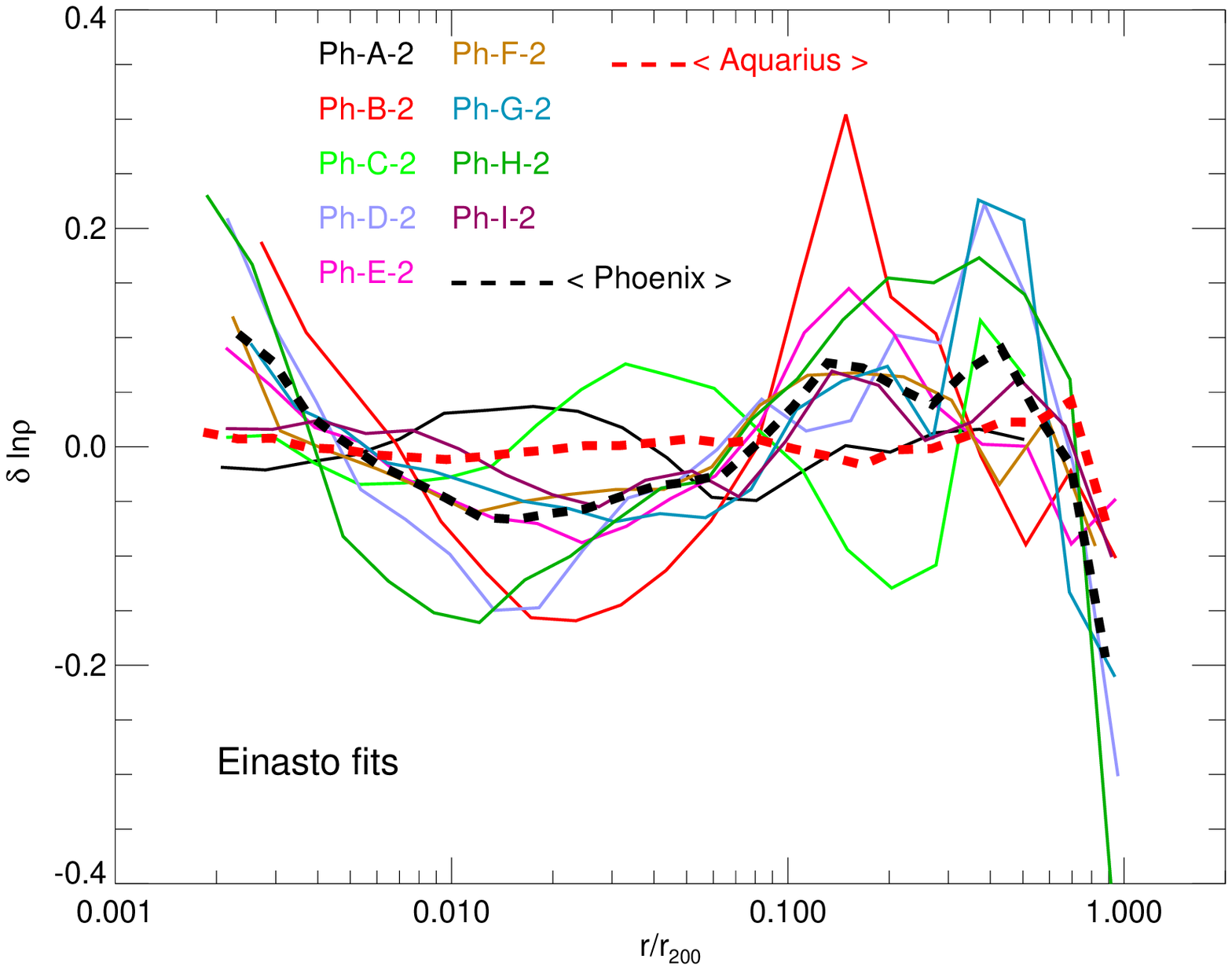}}
\hspace{0.13cm}\resizebox{8cm}{!}{\includegraphics{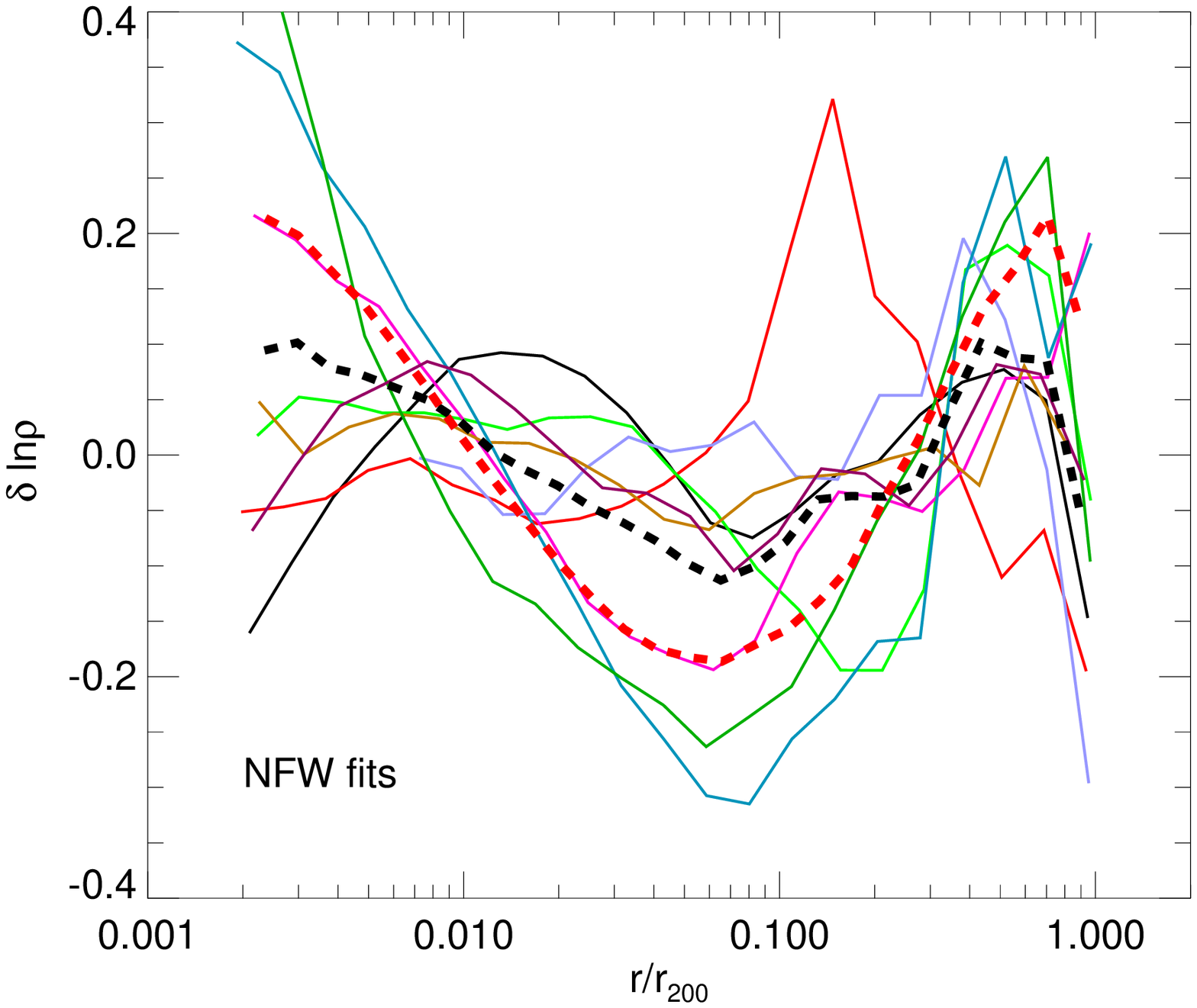}}
\caption{ Residuals from the best Einasto (left panel) and NFW (right
  panel) profile fits for all level-2 Phoenix haloes. Colors and line
  types are as in Fig.\ref{FigAllDensProf}. The thick black dashed curve
  corresponds to the composite profile obtained after stacking all 9
  Phoenix level-2 runs. The red thick dashed curve corresponds to the
  same composite profile, but for the 6 galaxy-sized level-2
  Aquarius haloes.}
\label{FigResid}
\end{figure*}

Fig.~\ref{FigAllDensProf} shows the density profiles of all level-2
Phoenix haloes, in a format similar to that of
Fig.~\ref{FigConvDensProf}. The top panels show profiles with radii
scaled to the virial radius of each cluster, whereas those at the
bottom show radii in units of the ``scale radius'', $r_{-2}$, of the
best Einasto fit. Profiles are shown from the convergence radius,
$r_{\rm conv}$, outwards.

In general, the density profiles of Phoenix clusters become gradually
shallower towards the centre; from $\gamma\sim 3$ in the outer regions
to an average value of $\gamma\sim 1$ at the innermost resolved
radius. This behaviour is similar to that of Aquarius haloes, whose
average profile is shown by the thick red dashed lines in
Fig.~\ref{FigAllDensProf}. The large difference between Aquarius and
Phoenix seen in the top panels of this figure just reflect the
different concentration\footnote{The concentration is defined as
  $r_{200}/r_{-2}$, where $r_{-2}$ is the radius at which the
  logarithmic slope $\gamma$ has
  the isothermal value of $2$. This indicates the location of the
  maximum of the curves shown in Fig.~\ref{FigAllDensProf}.} of
cluster- and galaxy-sized haloes. Indeed, when radii are scaled to
$r_{-2}$, {\it the average Phoenix and Aquarius profiles are basically
indistinguishable from each other}.

This is confirmed quantitatively by the best-fit Einasto parameters of
these average profiles (listed in Table~\ref{TabCompAqPh}). The
average Phoenix halo is only slightly worse fit by an Einasto profile
than Aquarius, as shown by the $Q_{\rm min}$ goodness-of-fit measure
($6.5\%$ vs $1.8\%$, respectively). There is also a slight difference
in shape parameter; the average Phoenix cluster has $\alpha=0.175$
whereas the average Aquarius halo has $\alpha=0.159$, in agreement
with previously reported trends \citep{Gao08}.

One aspect in which Phoenix and Aquarius haloes do differ is the
halo-to-halo scatter: the dispersion in the Einasto parameter $\alpha$
is twice as large for clusters as for galaxy-sized haloes
(Table~\ref{TabCompAqPh}). This may be readily seen in
Fig.~\ref{FigAllDensProf}: Ph-A-2, for example, follows the steady
decline in $\gamma$ towards the centre exhibited by Ph-A-1 (and
characteristic also of Aquarius haloes), whereas in other cases, such
as Ph-H-2, $\gamma$ stays roughly constant over a wide radial range
near the centre.

The latter behaviour is poorly captured by the Einasto or NFW fitting
formulae, and leads to larger residuals and figure-of-merit values for
the best fits. NFW and Einasto best-fit residuals are shown in
Fig.~\ref{FigResid}; per bin deviations of up to $40\%$ from NFW and
$\sim 20\%$ from Einasto fits are not uncommon for Phoenix clusters.

These deviations may be traced to transient departures from
equilibrium induced by the recent formation of many Phoenix
clusters. For example, one of the worst offenders is Ph-H-2, which
accreted half its final mass since $z=0.21$ and whose unrelaxed
appearance is obvious in Fig.~\ref{FigImAllClus}. In contrast, Ph-A-2,
the cluster with highest formation redshift of the Phoenix series
($z_{\rm h}=1.17$) is very well fit by both the Einasto and NFW
profiles, with average residuals of only $\sim 3\%$ and $\sim 6\%$,
respectively. Indeed, a well defined correlation may be seen in
Table~\ref{TabHaloParam} between quantitative measures of departures
from equilibrium, such as the centre offset, $d_{\rm off}$, or the
mass fraction in the form of substructure, $f_{\rm sub}$, and the
average residuals from the best NFW and Einasto fits. On average, both
indicators are substantially smaller for Aquarius than for Phoenix
(Table~\ref{TabCompAqPh}), as expected. The higher formation redshift
of galaxy-sized haloes means that they are closer to dynamical
equilibrium than recently-assembled cluster haloes.

\subsection{Pseudo-Phase-Space Density and Velocity Anisotropy }
\label{SecPPSD}

The similarity in the mass profiles of galaxy- and cluster-sized CDM
haloes highlighted in the previous subsection extends to their
dynamical properties. We show this by comparing the
spherically averaged pseudo-phase-space density (PPSD) profiles of
Phoenix and Aquarius haloes. The PPSD, $\rho/\sigma^3$, is
dimensionally identical to the phase-space density, but not strictly a
measure of it. (Here the velocity dispersion, $\sigma(r)$, is defined as
the square root of twice the specific kinetic energy in each
spherical shell.) It is well known that PPSD profiles are well
approximated by a simple power-law, $\rho/\sigma^3 \propto r^\chi$
\citep{tn01}, intriguingly similar to the secondary-infall
self-similar solutions of \citet{b85}, where the exponent,
$\chi=-1.875$ \citep[see also][and references therein]{ludlow10}.

\begin{figure}
\hspace{0.13cm}\resizebox{8cm}{!}{\includegraphics{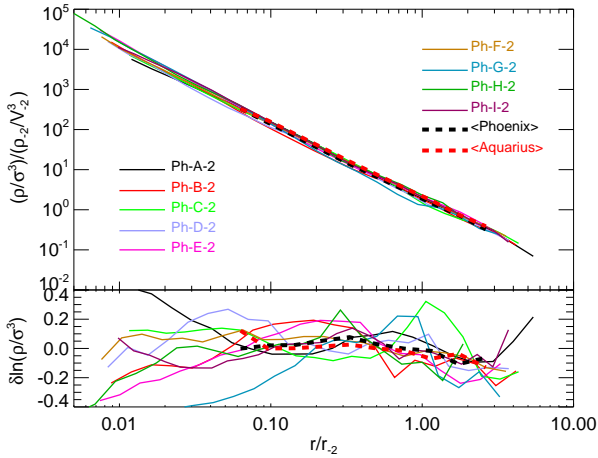}}
\caption{Spherically averaged pseudo-phase-space density (PPSD;
  $\rho/\sigma^3$) of all level-2 Phoenix haloes as a function of
  radius.  Profiles are plotted down to the convergence radius,
  $r_{\rm conv}$. Radii are given in units of the scale radius,
  $r_{-2}$, of the best-fit Einasto profile for each halo. Densities
  are scaled to $\rho_{-2}=\rho(r_{-2})$, and velocity dispersions,
  $\sigma(r)$, to the characteristic velocity
  $V_{-2}=(G\rho_{-2})^{1/2}\, r_{-2}$.  The thick dashed lines shows
  the average {\small PPSD} of all Phoenix (black) and Aquarius (red)
  haloes plotted over the converged radial range common to both
  simulation series: $0.06 \leq (r/r_{-2}) \leq 4$,
  respectively. The bottom panel shows residuals from a simple
  $r^{-1.875}$ power-law fit.}
\label{FigPPSD}
\end{figure}

PPSD profiles for all level-2 Phoenix clusters are shown in
Fig.~\ref{FigPPSD}, and compared with the average PPSD for Aquarius
haloes. Since clusters are denser and have higher velocity dispersions
than galaxy-sized haloes, we scale all profiles to the scale radius,
$r_{-2}$, of each halo. Together with the density at this
characteristic radius, $\rho_{-2}$, these quantities define a
characteristic velocity, $V_{-2}=(G\, \rho_{-2})^{1/2}\, r_{-2}$, that
allows us to compare PPSD profiles of haloes of widely different mass
in a meaningful way.

The top panel of Fig.~\ref{FigPPSD} shows that, in these scaled units,
Aquarius and Phoenix have very similar PPSD profiles. The similarity
extends over the range $0.06<r/r_{-2}<4$ where both simulation
sets give converged results. (Note that Phoenix profiles actually
probe radii interior to $0.06\, r_{-2}$ because of their lower
concentration.) Table~\ref{TabCompAqPh} lists the average parameters
(and their dispersion) of power-law fits of the form
\begin{equation}
{\rho \over \sigma^3} = C_\chi\, {\rho_{-2} \over V_{-2}^3} \, \biggl({r \over r_{-2}}\biggr)^{\chi},
\label{EqPPSD}
\end{equation}
where $C_\chi=(\sigma(r_{-2})/V_{-2})^3$. Fits are carried out over
the length range $0.06<r/r_{-2}<4$ for each halo. On average, both the
slope and the normalisation of Aquarius and Phoenix PPSD profiles are
almost indistinguishable emphasizing again the structural similarity
between cluster- and galaxy-sized haloes.  

At the same time, the scatter is larger in the Phoenix sample than in
Aquarius (Table~\ref{TabCompAqPh}), highlighting again the larger
halo-to-halo variation of cluster profiles. This may also be
appreciated in the bottom panel of Fig.~\ref{FigPPSD}, where residuals
from the self-similar $r^{-1.875}$ power law are shown. Although PPSD
profiles scatter above and below the self-similar solution depending
on the individual dynamical state of each cluster, the PPSD profile of
cluster haloes seem to be, on average, indistinguishable from that of
galaxy-sized haloes.

\begin{figure}
\hspace{0.13cm}\resizebox{8cm}{!}{\includegraphics{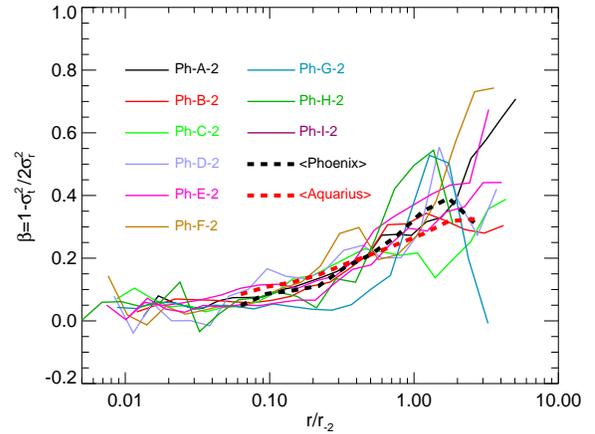}}
\caption{Velocity anisotropy profiles of all level-2 Phoenix haloes. Radii are
  expressed in units of the scale radius, $r_{-2}$, of the best-fit
  Einasto profile. Profiles are plotted down to the convergence
  radius, $r_{\rm conv}$. The thick dashed lines show the average
  anisotropy profile of all Phoenix (black) and Aquarius (red) haloes 
  over the radial range where both give converged results $(0.06\leq
  (r/r_{-2}) \leq 4)$.}
\label{FigVelAnis}
\end{figure}

We reach a similar conclusion when comparing the velocity anisotropy
profiles of Phoenix clusters with those of Aquarius haloes
(Fig.~\ref{FigVelAnis}). Aside from a slightly larger scatter, the
velocity anisotropy, which measures the ratio of the kinetic energy in
tangential and radial motions, increases gently from the centre, where
haloes are nearly isotropic, to the outer regions, where radial
motions dominate. Phoenix and Aquarius haloes again seem
indistinguishable from each other regarding velocity anisotropy when
compared over their converged radial range.

\subsection{Projected Profiles}
\label{SecProjProf}

The preceding discussion highlights the mass invariance of the
structure of CDM haloes, but it also makes clear that the dynamical
youth of clusters limits the validity of simple fitting formulae to
describe their instantaneous mass profiles. This complication must be
taken into account when comparing observational estimates of
individual cluster mass profiles with the profiles expected in a
CDM-dominated Universe.  Stacking clusters in order to obtain an
``average'' cluster profile might offer a way of circumventing this
difficulty. This should smooth out local inhomogeneities in the mass
distribution and average over different dynamical states to produce a
more robust measure of the shape of the mass profile.

Aside from dynamical youth, another issue that complicates the
interpretation of observations is the fact that, due to the cluster's
asphericity, {\it projected} mass profiles, such as those measured
through gravitational lensing, may differ substantially from simple
projections of the 3D spherically-averaged profiles discussed above.

\begin{figure*}
\hspace{0.13cm}\resizebox{8cm}{!}{\includegraphics{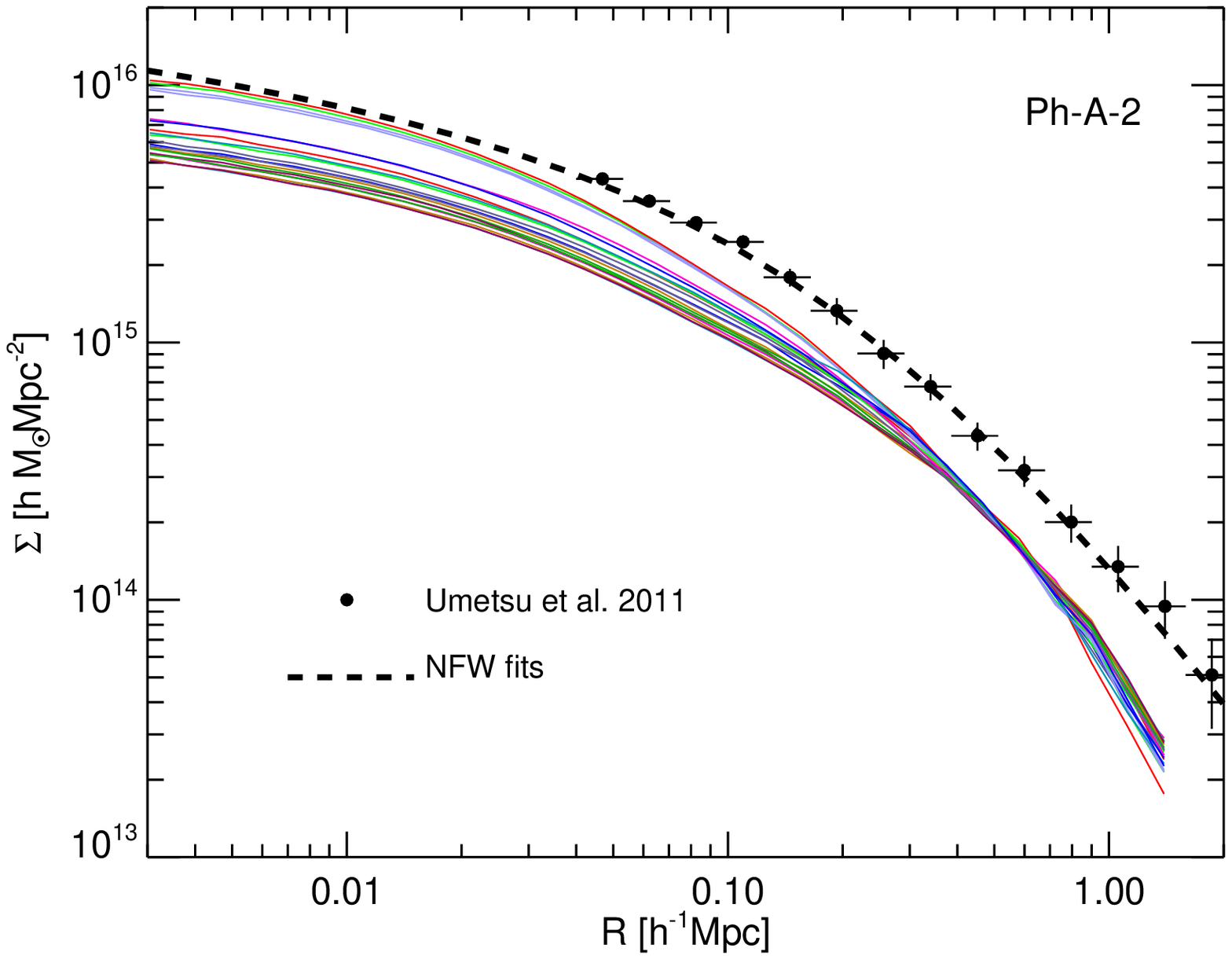}}
\hspace{0.13cm}\resizebox{8cm}{!}{\includegraphics{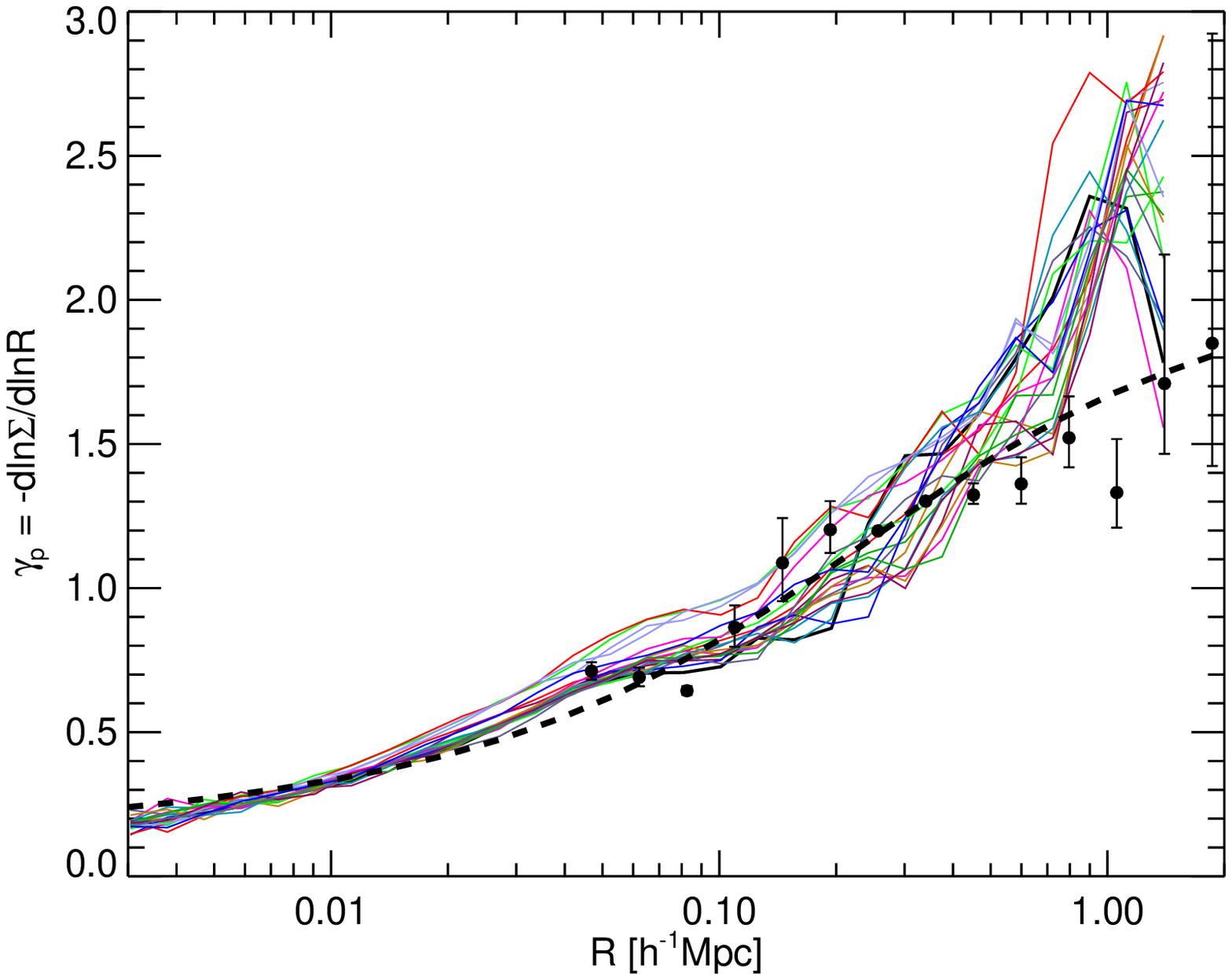}}
\hspace{0.13cm}\resizebox{8cm}{!}{\includegraphics{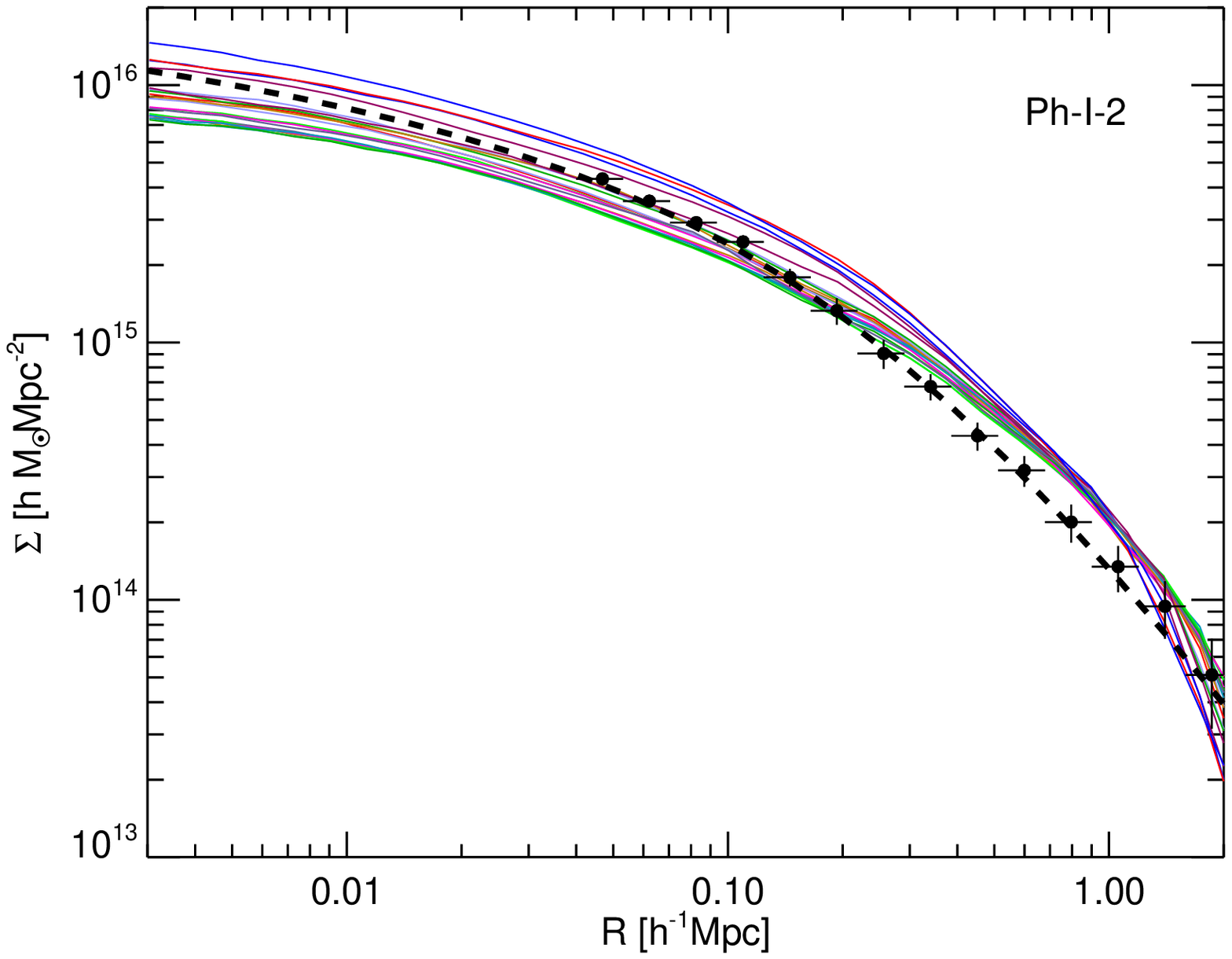}}
\hspace{0.13cm}\resizebox{8cm}{!}{\includegraphics{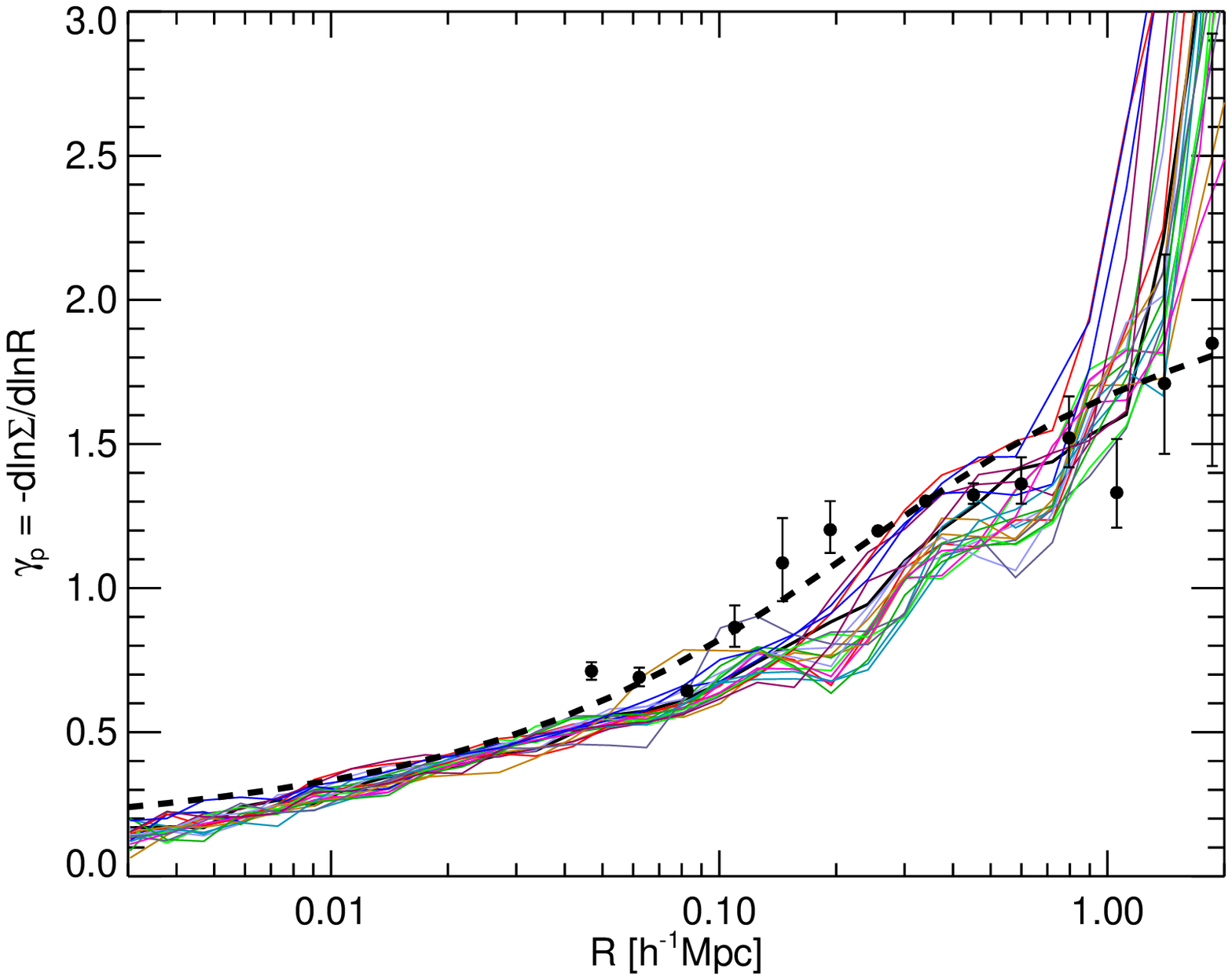}}
\caption{Projected density profiles of Ph-A-2 (top) and Ph-I-2
  (bottom). We show 20 different random projections for each cluster.
  The asphericity of the clusters leads to large variations (up to a
  factor of 3) in the projected density at a given radius depending on
  the line of sight. On the other hand, the {\it shape} of the profile
  (as measured by the logarithmic slope, $\gamma_p=-d\ln \Sigma/d \ln
  R$, is much less sensitive to projection effects. Data with error
  bars correspond to the stacked profile of 4 massive clusters
  estimated using strong and weak lensing data \citep{Umetsu2011}.}
\label{FigProjProf} 
\end{figure*}

Depending on the line of sight a cluster may appear more or less
massive within a given radius, leading to biases in the cluster's
estimated mass, concentration, and even the shape of its density
profile. We illustrate this in Fig.~\ref{FigProjProf}, where we plot the
surface density profile of two Phoenix clusters, Ph-A-2 and Ph-I-2,
each projected along $20$ different random lines of sight. The
aspherical nature of the clusters results in large variations (up to a
factor of 3) in the surface density in the inner regions. For
comparison, we also show in Fig.~\ref{FigProjProf} the result of a
weak and strong-lensing analysis of a stack of four massive clusters
by \citet{Umetsu2011}.  The mass of the stacked cluster lies
between that of Ph-A and Ph-I, which explains why, on average, Ph-A
$\Sigma(R)$ profiles lie below the observed data whereas the opposite
applies to Ph-I.

In agreement with earlier work \citep[see, e.g.,][and references
therein]{Corless2007,Sereno2010,Oguri2010}, Fig.~\ref{FigProjProf}
suggests that substantial biases may be introduced by projection
effects on estimates of cluster parameters, especially when reliable
data are restricted to the inner regions of a cluster. For example,
fitting the inner $500 \, h^{-1}$ kpc of the Ph-A-2 projected profile
with an NFW profile results in mass and concentration ($M_{200}$, $c$)
estimates that vary from ($5.4\times 10^{14} \, h^{-1} \, M_\odot$,
$4.8$) to ($7.3 \times 10^{14} \, h^{-1} \, M_\odot$, $9.8$) when
using the projections that maximize or minimize the inner surface
density, respectively (see Fig.~\ref{FigProjProf}). The corresponding
numbers for Ph-I-2 are ($1.8 \times 10^{15} \, h^{-1} \, M_\odot$,
$4.1$) and ($3.0 \times 10^{15}\, h^{-1} \,M_\odot$, $7.1$). Comparing
these numbers with those listed in Tables~\ref{TabSimParam}
and~\ref{TabHaloParam} we see that variations as large as $\sim 30\%$
in the mass and $\sim 60\%$ in the concentration may be introduced
just by projection effects\footnote{Note that variations may actually
  be larger, because these estimates neglect the possible contribution
  of the large-scale mass distribution along the line-of-sight.}.

\begin{figure}
  \hspace{0.13cm}\resizebox{8cm}{!}{\includegraphics{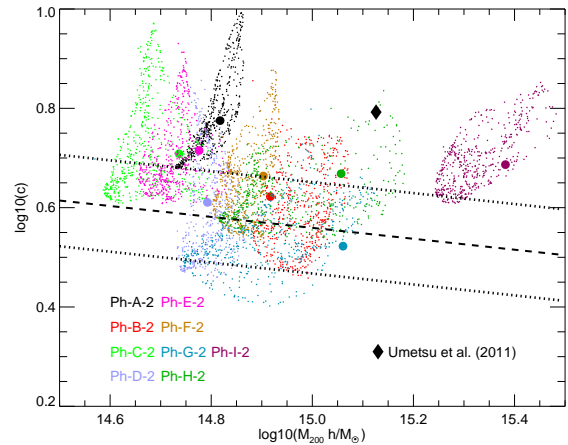}}
  \caption{Cluster virial mass vs concentration estimated from fits to
    the projected density profiles of level-2 Phoenix haloes in the
    radial range $R<500\, h^{-1}$ kpc. A total of $500$ random
    projections are used for each halo. The large filled circles
    indicate the true value of the virial mass and concentration of
    the cluster, obtained from NFW fits to the 3D spherically-averaged
    profile (see Appendix and Table~\ref{TabHaloParam}).  The dashed
    curve flanked by dotted lines shows the fit to the
    mass-concentration relation derived by \citet{Neto07}. Note that
    projection effects lead to significant bias in the mass and
    concentration.  , which are underestimated on average by $8.5 \pm
    17 \%$ and $0.4 \pm 20\%$, respectively, where the ``error'' is
    the rms of all projections for the 9 clusters.  The black diamond
    symbol indicates the $M_{200}$-$c$ estimate for a stack of 4
    strong-lensing clusters taken from \citet{Umetsu2011}.}
  \label{FigMassConc} 
\end{figure}

\begin{figure}
  \hspace{0.13cm}\resizebox{8cm}{!}{\includegraphics{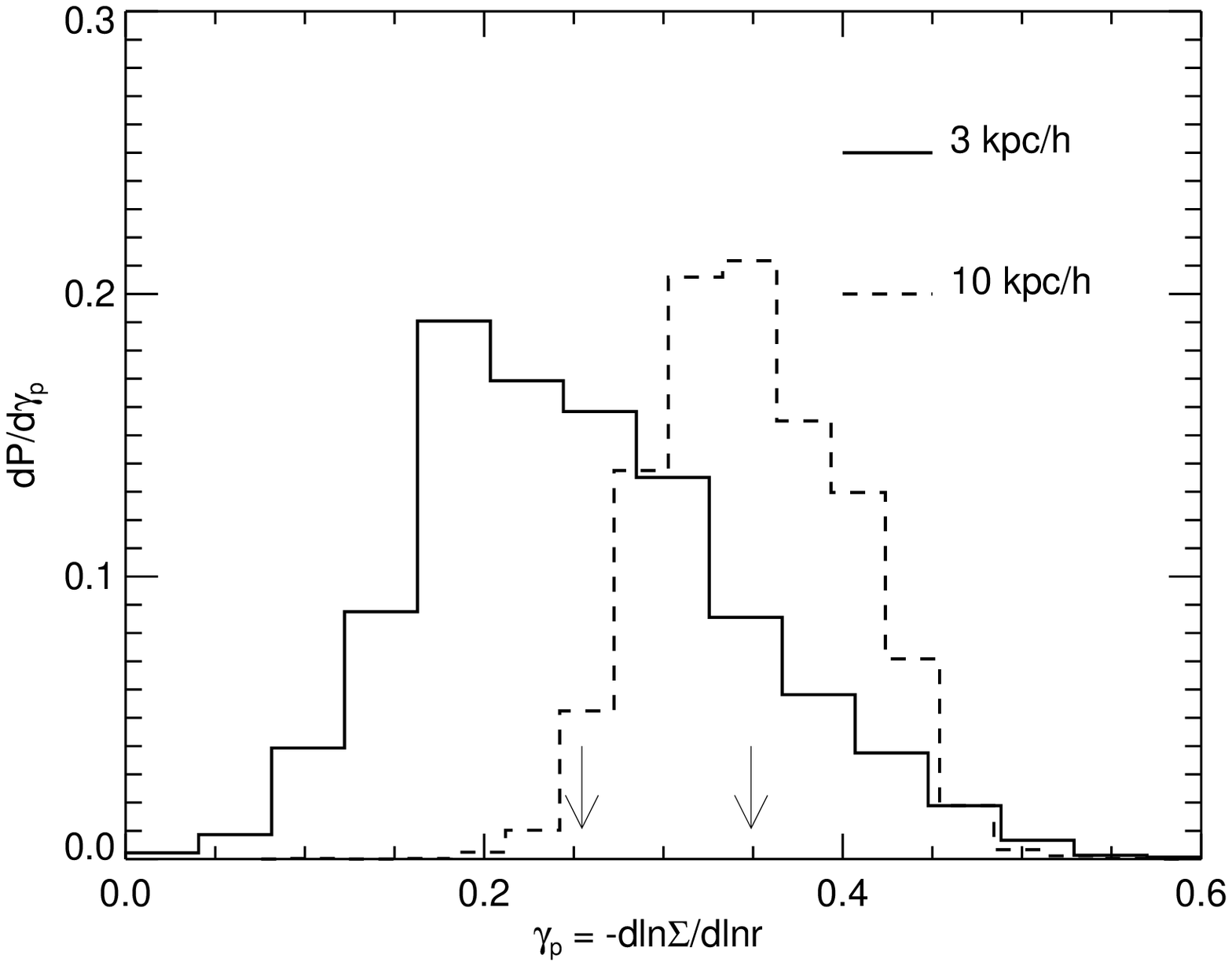}}
  \caption{Distribution of the slope of the circularly-averaged
    surface density profile, $\gamma_p(R)$, measured at two different
    radii, $R=3$ and $10 \, h^{-1}$ kpc in projection. These
    histograms are based on $500$ random lines of sight for each of
    the level-2 Phoenix clusters. Vertical arrows show the values
    corresponding to the projected profile of all nine clusters
    stacked together. The profiles become gradually shallower towards
    the centre, but with large scatter: $\langle \gamma_p \rangle$
    changes from $0.35$ to $0.25$ as $R$ goes from 10 to $3 \, h^{-1}$
    kpc, but the halo-to-halo scatter is quite large, with rms of
    order $0.09$ at $3 \, h^{-1}$ kpc and $0.05$ at $10 \, h^{-1}$
    kpc, respectively.}  
\label{FigSlopeHist} 
\end{figure}

We explore this further in
Fig.~\ref{FigMassConc}, where the small dots show the
mass-concentration estimates for 500 random projections of each
level-2 Phoenix cluster.  Large symbols correspond to the 3D estimates
listed in Tables~\ref{TabSimParam} and~\ref{TabHaloParam}. The black
diamond symbol indicates the $M_{200}$-$c$ estimate for the stack of 4
strong-lensing clusters presented by \citet{Umetsu2011}. This figure
again emphasizes the importance of projection effects; for example,
$12\%$ of random projections result in concentration overestimates
larger than $25\%$.  Although an exhaustive analysis of such biases is
beyond the scope of the present paper, the results in
Figs.~\ref{FigProjProf} and ~\ref{FigMassConc} suggest that there is
no substantial difficulty matching the surface density profile of
lensing clusters such as those studied by \citet{Umetsu2011}. Our
interpretation thus agrees with that reached by a number of recent
studies \citep[see,
e.g.,][]{Oguri2011,Okabe2010,Gralla2011,Umetsu2011}, which conclude
that there is no obvious conflict between the concentration of
lensing-selected clusters and those of $\Lambda$CDM haloes once
projection effects are taken into account.  Interestingly, despite the
large variations in surface density alluded to above, the {\it shape}
of the surface density profile is quite insensitive to projection
effects. We show this in the right-hand panels of
Fig.~\ref{FigProjProf}; the weak dependence of $\gamma_p(R)$ on
projection may thus be profitably used to assess the consistency of
theoretical predictions with cluster mass profiles.  For illustration,
we compare in the same panels the logarithmic slope of the projected
profile, $\gamma_p=d\ln \Sigma(R)/d\ln R$, with the stacked cluster
data of \citet{Umetsu2011}. Despite the fact that the mass of the
simulated and observed clusters are different and that no scaling has
been applied, there is clearly quite good agreement between
observation and Phoenix clusters, supporting our earlier conclusion.
Available data on individual clusters are bound to improve
dramatically with the advent of surveys such as CLASH with the
Advanced Camera for Surveys onboard the Hubble Space Telescope
\citep{postman11}. These surveys will enable better constraints on the
shape of the inner mass profile of individual rich clusters and it is
therefore important to constrain how projection effects may affect
them. Fig.~\ref{FigSlopeHist} shows the distribution of $\gamma_p$ at
two projected radii, $R=3$ and $10 \, h^{-1}$ kpc. The histograms are
computed after choosing $500$ random lines of sight for each of our 9
level-2 Phoenix haloes. On average, cluster projected profiles flatten
steadily towards the centre, from $\langle \gamma_p \rangle = 0.35$ to
$0.25$ in that radial range, but with fairly large dispersion; the rms
is $\sigma_{\gamma_p}=0.054$ and $0.091$ at $R=10$ and $3\, h^{-1}$
kpc respectively. Because of the large dispersion it is unlikely that
observations of a single cluster can lead to conclusive statements
about the viability of CDM; however, it should be possible to use this
constraint fruitfully once data for a statistically significant number
of clusters become available.   

\section{The Substructure of Phoenix Clusters} \label{SecSubs} 

As may be seen from the images presented in Fig.~\ref{FigImAllClus},
substructure is ubiquitous in Phoenix clusters. We have used {\small
  SUBFIND} \citep{Springel01a} to identify and characterize self-bound
structures (subhaloes) within the virial radius of the main halo. We
discuss below the mass function, spatial distribution, and internal
properties of subhaloes in Phoenix. Since our main goal is to explore
the mass invariance of the properties of CDM haloes, we contrast these
results with those obtained for the galaxy-sized Aquarius haloes.
\subsection{Mass Function} \label{SecSubsMF} 
\begin{figure*}
  \hspace{0.13cm}\resizebox{8cm}{!}{\includegraphics{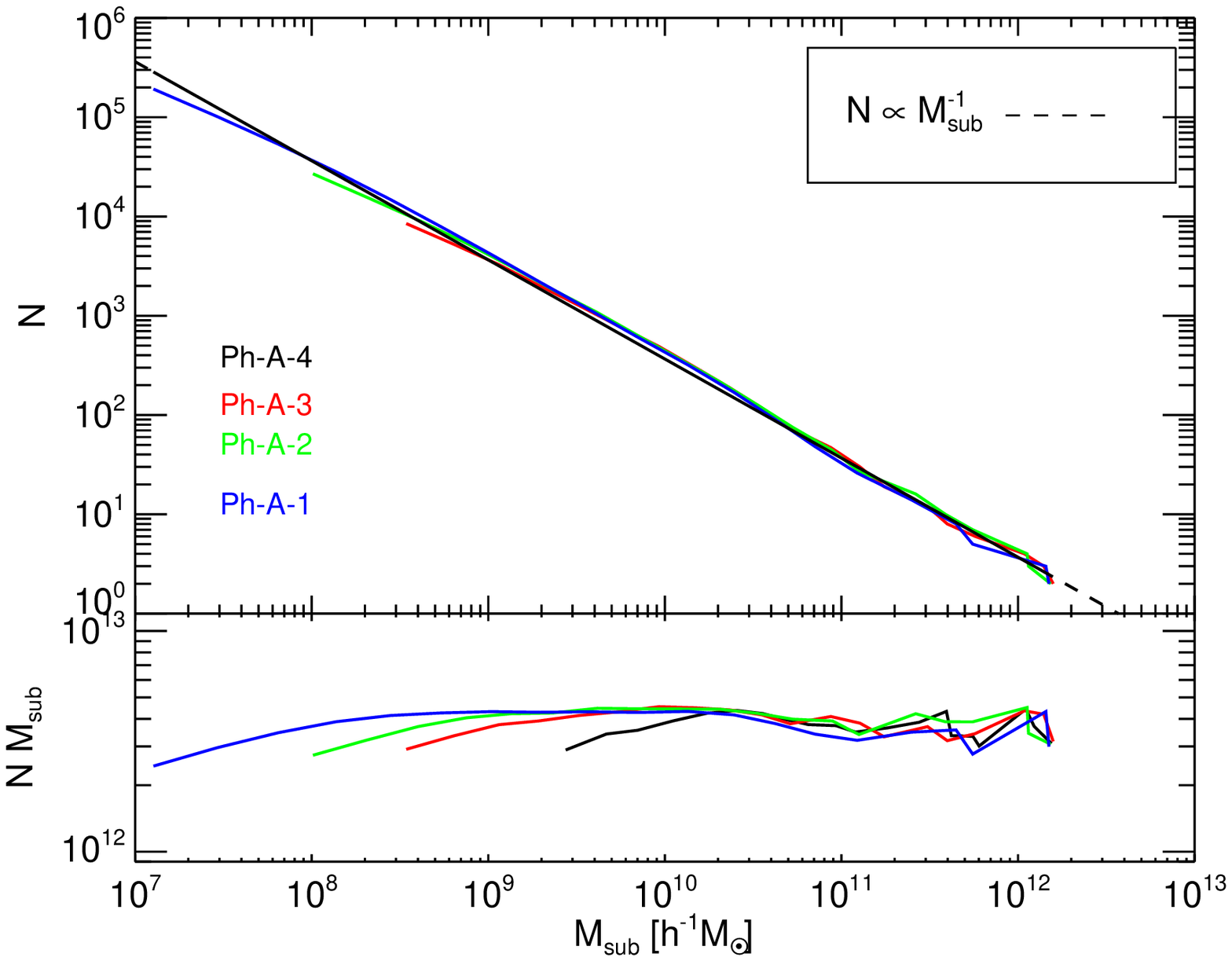}}
  \hspace{0.13cm}\resizebox{8cm}{!}{\includegraphics{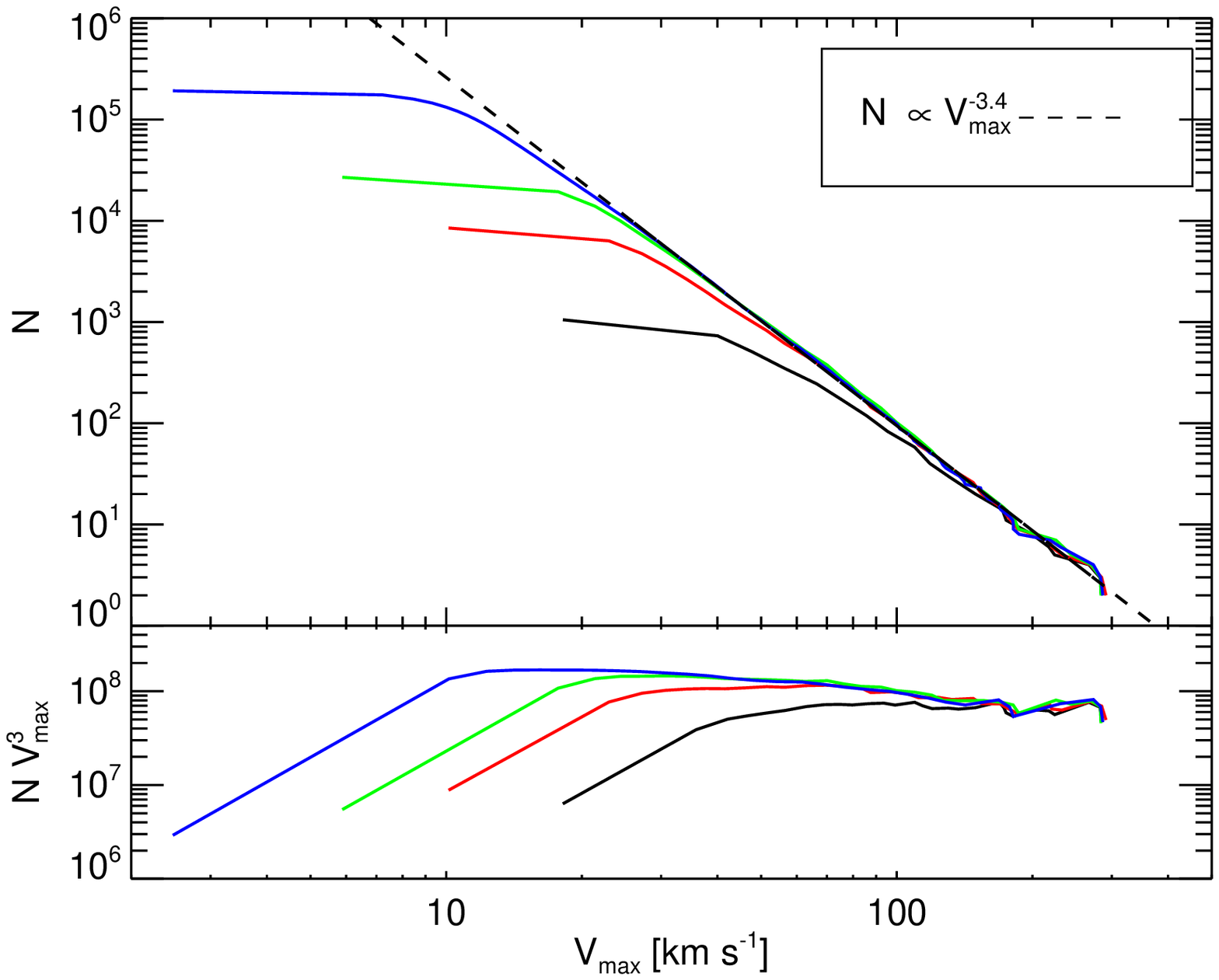}}
  \caption{{\it Left:} The cumulative mass function of substructure
    haloes (``subhaloes'') within the virial radius of cluster Ph-A at
    $z=0$. We compare the results of four different realizations of
    the same halo, Ph-A-1 to Ph-A-4, with varying numerical
    resolution. The top and bottom panels contain the same
    information; the bottom shows the number of subhaloes weighted by
    mass or, equivalently, the fractional contribution of each
    logarithmic mass bin to the total mass in subhaloes. Each curve
    extends down to a mass corresponding to $60$ particles. Note that,
    over the range resolved by the simulations, the cumulative
    function is well approximated by a power-law, $N\propto M^{-1}$,
    the critical dependence for logarithmically divergent substructure
    mass.  {\it Right:} Same as left panels, but for the subhalo peak
    circular velocity.}  
\label{FigPhASubsMF} 
\end{figure*}

We start by analyzing the
Ph-A simulation series in order to identify the limitations introduced
by finite numerical resolution. The top left panel of
Fig.~\ref{FigPhASubsMF} shows the cumulative mass function of
subhaloes, $N(>M)$, plotted in each case down to the mass
corresponding to $60$ particles. The bottom left panel shows the same
data, but after weighting the numbers by subhalo mass, $M_{\rm sub}$,
in order to emphasize the differences between runs.  The results show
clearly how, as resolution improves, the mass function converges at
the low-mass end. Ph-A-4 agrees with higher resolution runs for
subhaloes with mass exceeding $\sim 2\times 10^{10} \, h^{-1} \,
M_\odot$, corresponding to roughly $150$ particles; the same applies
to Ph-A-3 for mass greater than $\sim 3\times 10^{9} \, h^{-1} \,
M_\odot$, or $\sim 170$ particles, and to Ph-A-2 for $\sim 7\times
10^{8} \, h^{-1} \, M_\odot$, or $140$ particles. We conclude that the
subhalo mass function can be robustly determined in Phoenix haloes
down to subhaloes containing roughly $150$ particles, in good
agreement with the results reported for Aquarius haloes \citep[see
Fig. 6 of][]{sp08b}. For level-2 runs this implies a subhalo mass
function that spans over $6$ decades in mass below the virial mass of
the halo.  The subhalo mass function is also routinely expressed in
terms of the subhalo peak circular velocity. This is shown in the
right-hand panels of Fig.~\ref{FigPhASubsMF} which shows that level-2
Phoenix runs give robust estimates of the abundance of subhaloes down
to $V_{\rm max} \sim 20$ km s$^{-1}$, a factor of $\sim 75$ lower than
the main halo's $V_{200}$.  

\begin{figure*}
  \hspace{0.13cm}\resizebox{8cm}{!}{\includegraphics{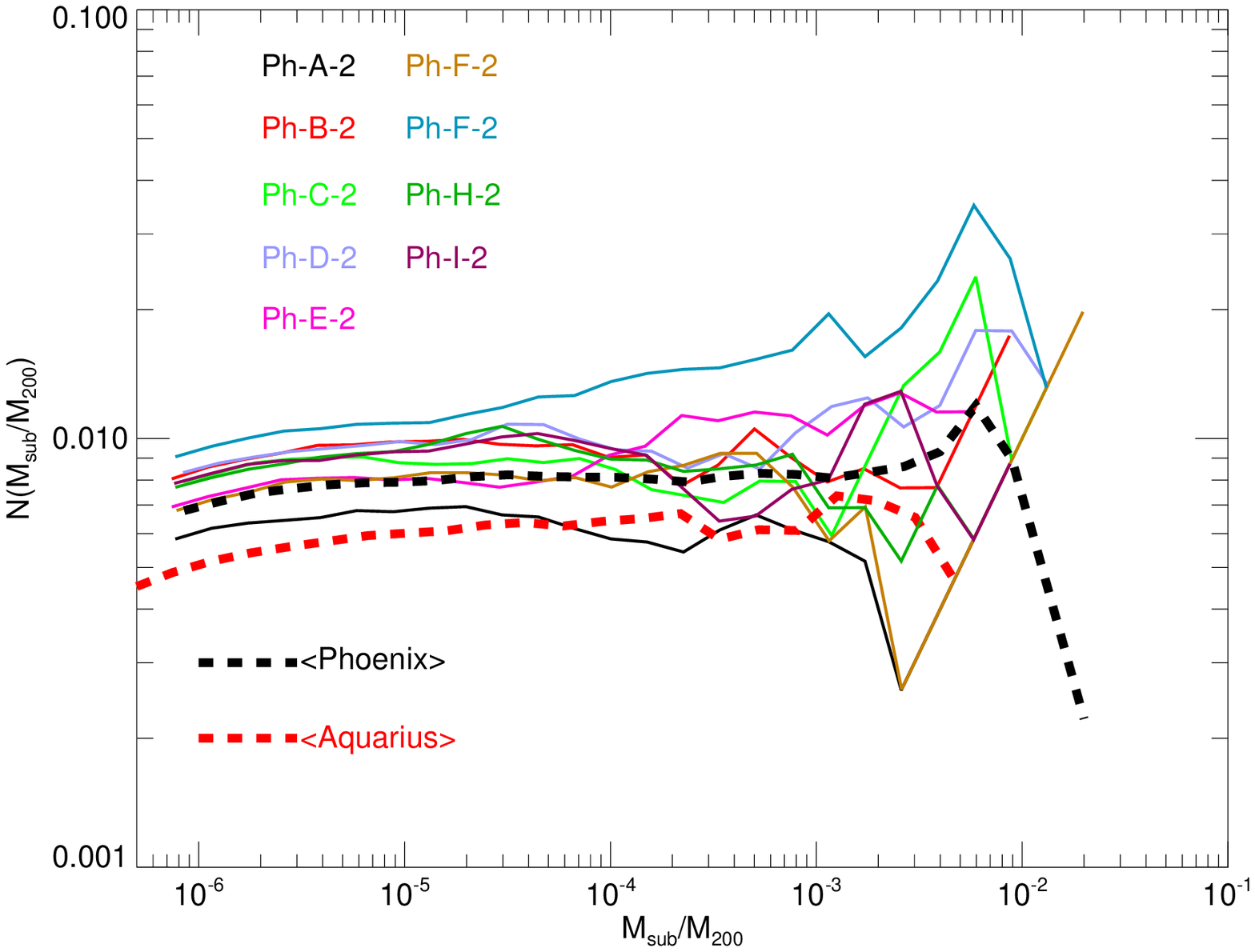}}
  \hspace{0.13cm}\resizebox{8cm}{!}{\includegraphics{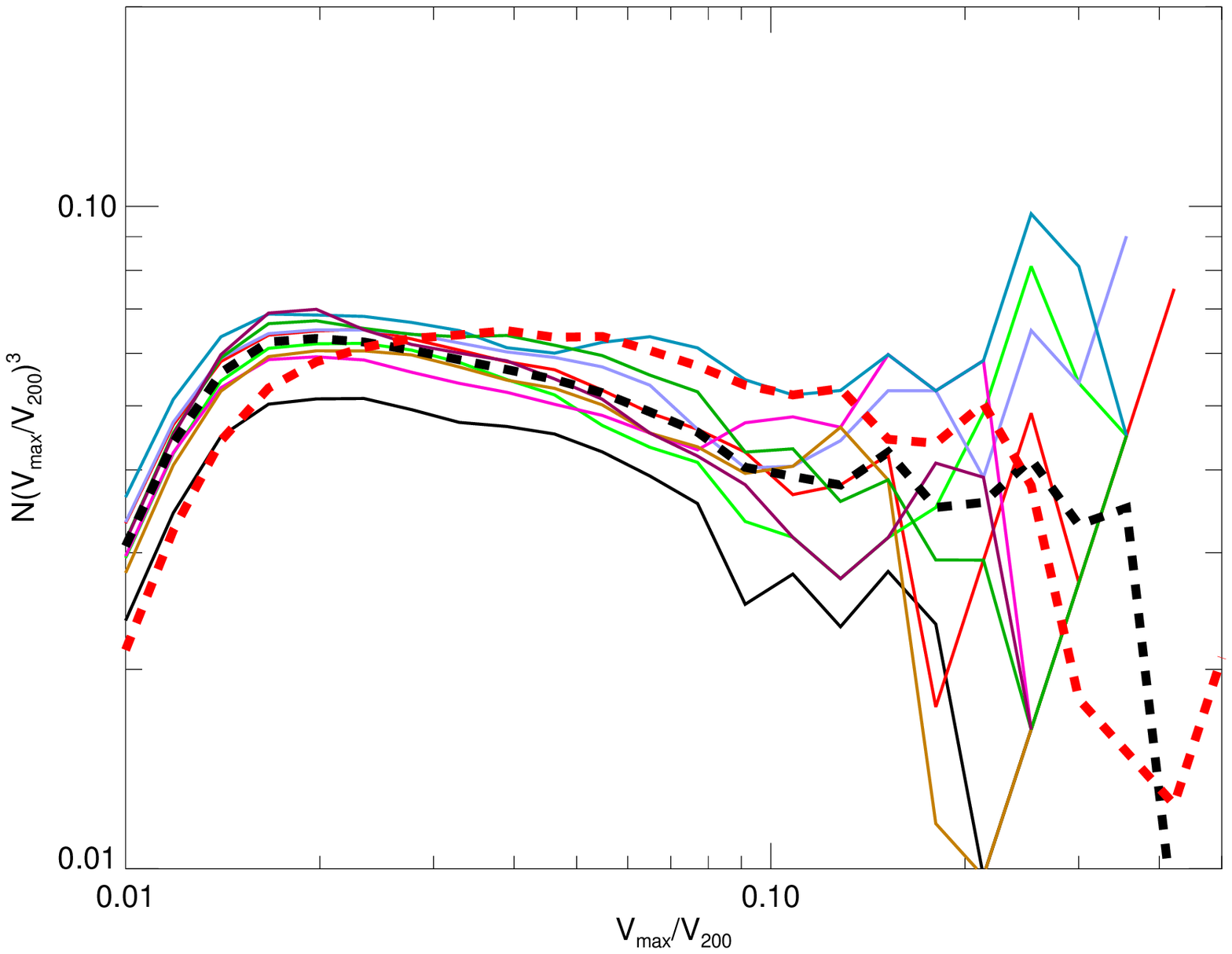}}
  \caption{As the bottom panels of Fig.~\ref{FigPhASubsMF}, but for
    all level-2 Phoenix haloes. The cumulative mass function (left
    panel) is weighted by subhalo mass, expressed in units of the
    virial mass. A cumulative $N\propto M^{-1}$ dependence, the
    critical case for logarithmic divergence in the total substructure
    mass, corresponds to a horizontal curve in these scaled units.
    Although the dependence is nearly flat in several Phoenix clusters
    it is clearly declining in others, and the average trend seems to
    be sub-critical. Compared with Aquarius (thick dashed red curve)
    the average Phoenix subhalo mass function is slightly steeper. The
    panel on the right is analogous to the mass function, but for the
    subhalo peak circular velocity, weighted by $V_{\rm max}^3$.  (See
    text for further discussion.)  } \label{FigAllSubhMF}
\end{figure*} 

Both the subhalo mass and velocity functions seem
reasonably well approximated by simple power laws: $N\propto M_{\rm
  sub}^{-1}$ and $N\propto V_{\rm max}^{-3.4}$, respectively.
Interestingly, the $M^{-1}$ dependence corresponds to the critical
case where each logarithmic mass bin contributes equally to the total
mass in substructure. This is logarithmically divergent as $M_{\rm
  sub}$ approaches zero, and implies that a significant fraction of
the mass could in principle be locked in haloes too small to be
resolved by our simulations. We note, however, that even at the
resolution of Ph-A-1, of nearly $7$ decades in mass, only $8\%$ of the
mass within $r_{200}$ is in the form of substructure. Extrapolating
down to the Earth mass by assuming that $N\propto M_{\rm sub}^{-1}$,
the total mass locked in substructure would still be only about $27$
percent.  

Fig.~\ref{FigAllSubhMF} compares these results with other level-2
Phoenix clusters in order to assess the general applicability of the
Ph-A subhalo mass function. The cumulative number of subhaloes $N(>M)$
is weighted here by $\mu=M_{\rm sub}/M_{200}$ (left panel) in order to
emphasize differences as well as to enable the comparison of haloes of
different virial mass. Although the subhalo mass function, expressed
in this form, is relatively flat in several Phoenix clusters
(indicative of an $N\propto M_{\rm sub}^{-1}$ dependence) it is
clearly declining in others. The average trend, as indicated by the
``stacked'' Phoenix cluster (thick dashed black curve) may be
approximated, in the range $10^{-6}<\mu<10^{-4}$, by $N\propto
\mu^{-0.98}$. This is a slightly steeper dependence than found for
Aquarius haloes {\it over the same mass range}, $N\propto \mu^{-0.94}$
(thick dashed red curve), but still subcritical.  The slight
difference in the average slope of the Aquarius and Phoenix subhalo
mass functions is smaller than the halo-to-halo scatter in either
simulation set. This is shown in Table~\ref{TabCompAqPh}, where we
list the average parameters of power-law fits of the form,
\begin{equation} N(>\mu)=N_{m} \, (\mu/10^{-6})^{s} \label{EqSubMF},
\end{equation} for Aquarius and Phoenix haloes. The dispersion around
$\langle s \rangle$ is similar to the difference between the average
slope of Aquarius and Phoenix haloes, suggesting that there is no
significant difference in the {\it shape} of the subhalo mass function
of cluster- and Milky-way halo-sized haloes.  
\begin{figure}
  \hspace{0.13cm}\resizebox{8cm}{!}{\includegraphics{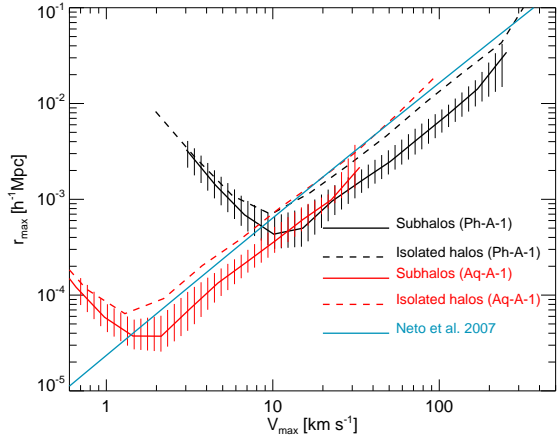}}
  \caption{Peak circular velocity, $V_{\rm max}$, vs the radius at
    which it is reached, $r_{\rm max}$. The solid cyan curve indicates
    the $r_{\rm max}$-$V_{\rm max}$ relation obtained for isolated
    haloes in the Millennium Simulation by \citet{Neto07}. Subhaloes
    in both Phoenix (solid black curve) and Aquarius (solid red curve)
    deviate systematically from this relation towards smaller $r_{\rm
      max}$ at a given velocity. This is a result of tidal stripping,
    which shifts the location of the peak inwards while changing
    little the peak velocity. Isolated haloes identified in Aquarius
    and Phoenix (shown with dashed lines) are not subject to tides and
    are in good agreement with the Millennium Simulation results. }
  \label{FigRmaxVmax} 
\end{figure} 

Fig.~\ref{FigAllSubhMF} also shows
that substructure is slightly more prevalent in clusters than in
galaxy-sized haloes. Indeed, at all values of $M_{\rm sub}/M_{200}$
the number of Phoenix subhaloes exceeds that in Aquarius, and this is
reflected in the higher values of $\langle N_{m} \rangle$ ($7866$ for
Phoenix vs $5092$ for Aquarius; see Table~\ref{TabCompAqPh}).  This is
another consequence of the dynamical youth of clusters compared to
galaxies (tides take a few orbital times to strip a subhalo), as may
be verified by inspection of Table~\ref{TabSimParam}: in the cluster
that forms latest, Ph-G, substructure makes up roughly $17\%$ of its
virial mass, but in the earliest collapsing system of the Phoenix
series, Ph-A, it makes only $8\%$.

Interestingly, as a function of $\nu=V_{\rm max}/V_{200}$, the comparison
between the Aquarius and Phoenix subhalo functions reverses
(right-hand panel of Fig.~\ref{FigAllSubhMF}). At a given velocity
(scaled to the virial value), subhaloes are more abundant in Aquarius than in
Phoenix. This is a consequence of tidal stripping, which affects
Aquarius subhaloes more: since tides act to remove
preferentially the outer regions of a subhalo, they affect more its
mass than its peak circular velocity.

For example, as discussed by \citet{Penarrubia08}, after losing {\it
half} of its mass to tides, the peak velocity of a subhalo decreases
only by $\sim 25\%$. Even after losing $90\%$ of its mass, $V_{\rm
max}$ is only reduced by about one half. Since Aquarius haloes
form earlier, their subhaloes have been accreted earlier and, on average, have
been more stripped than Phoenix subhalos,
leading to higher relative velocities for their bound mass than for
Phoenix subhalos. This shifts their abundance when measured in terms of peak velocity. In
the range $0.025<\nu<0.1$ fits to the subhalo function of the form
\begin{equation}
N(>\nu)=N_v\, (\nu/0.025)^d \label{EqSubVF}
\end{equation}
yield $\langle N_v \rangle=4033$ and $\langle d \rangle=-3.13$ for Aquarius and $3984$
and $-3.32$, respectively, for Phoenix (see
Table~\ref{TabCompAqPh}). Given the scatter, the difference seems too
small to be significant. We conclude that the scaled subhalo velocity
function, $N(>\nu)$, is roughly independent of mass \citep[see][for a
more thorough discussion of this point]{Wang2012}.

The effects of tidal stripping on Phoenix subhaloes is shown in
Fig.~\ref{FigRmaxVmax}. Here we plot $V_{\rm max}$ vs $r_{\rm max}$
for subhaloes identified in Ph-A-1 (solid black curve). This relation
is clearly offset from the mean relation that holds for isolated haloes
in the Millennium Simulation, as given by \citet{Neto07} (cyan
line). As expected for haloes that have undergone tidal stripping,
$r_{\rm max}$ shifts inwards as the subhalo loses mass whilst leaving
the peak velocity relatively unchanged \citep{Penarrubia08}. Support
for this interpretation may be found by inspecting the same relation
for ``isolated'' haloes in Phoenix (i.e., those outside the main halo
and that are not embedded in a more massive structure; the
$r_{\max}$-$V_{\rm max}$ relation for these systems (see dashed lines) is
consistent with that of Millennium haloes.

Fig.~\ref{FigRmaxVmax} also includes results for isolated haloes and
subhaloes in Aquarius (red lines). The results from the two sets of
simulations form a single sequence and this allows us to characterize the structural
parameters of subhaloes over a range spanning more than two decades in
velocity (and thus over six decades in mass). On average, subhaloes
follow the same $r_{\rm max}$-$V_{\rm max}$ scaling relations as
isolated haloes, but shifted by about a factor of two in radius (or,
alternatively, by $\sim 30\%$ in velocity).

We conclude from this discussion that although substructure does
not seem fully invariant with halo mass, the changes are relatively
small when comparing the haloes of clusters and galaxies, and depend on
whether subhalo masses or velocities are used to characterize
substructure. The subhalo mass function of clusters, scaled to halo
virial mass, is similar in shape to that of galaxy-sized haloes (which
are roughly one thousand times less massive), but with a slightly
higher normalization ($\sim 35\%$). The normalization difference
disappears when the scaled subhalo velocity function, $N(>\nu)$, is
used. The total mass in substructure increases with the dynamical
youth of the system and is more prevalent in clusters than on galaxy
scales, but only weakly so: the average mass fraction in substructures
is $11\%$ for Phoenix and $7\%$ for Aquarius.

\begin{figure*}
\hspace{0.13cm}\resizebox{8cm}{!}{\includegraphics{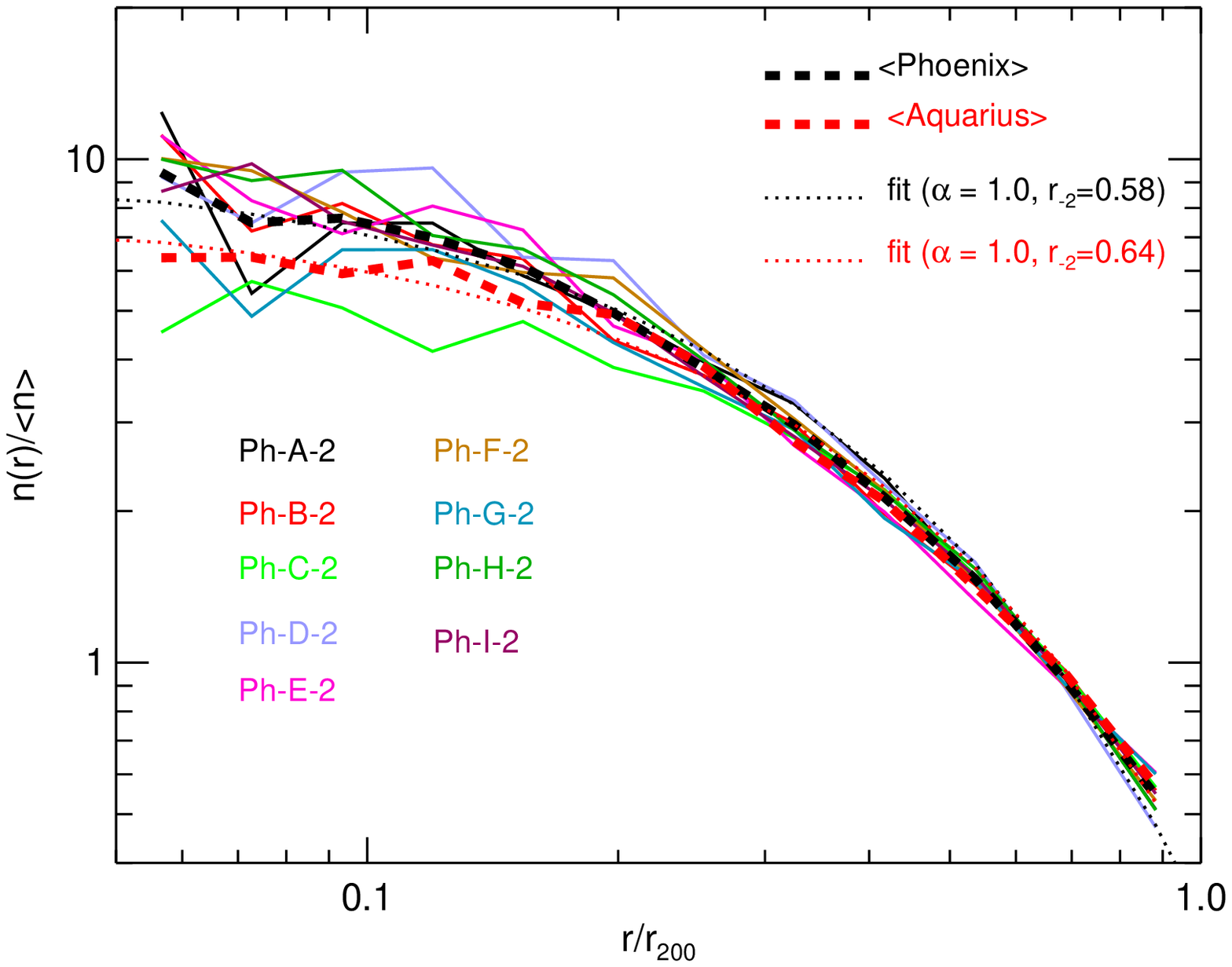}}
\hspace{0.13cm}\resizebox{8cm}{!}{\includegraphics{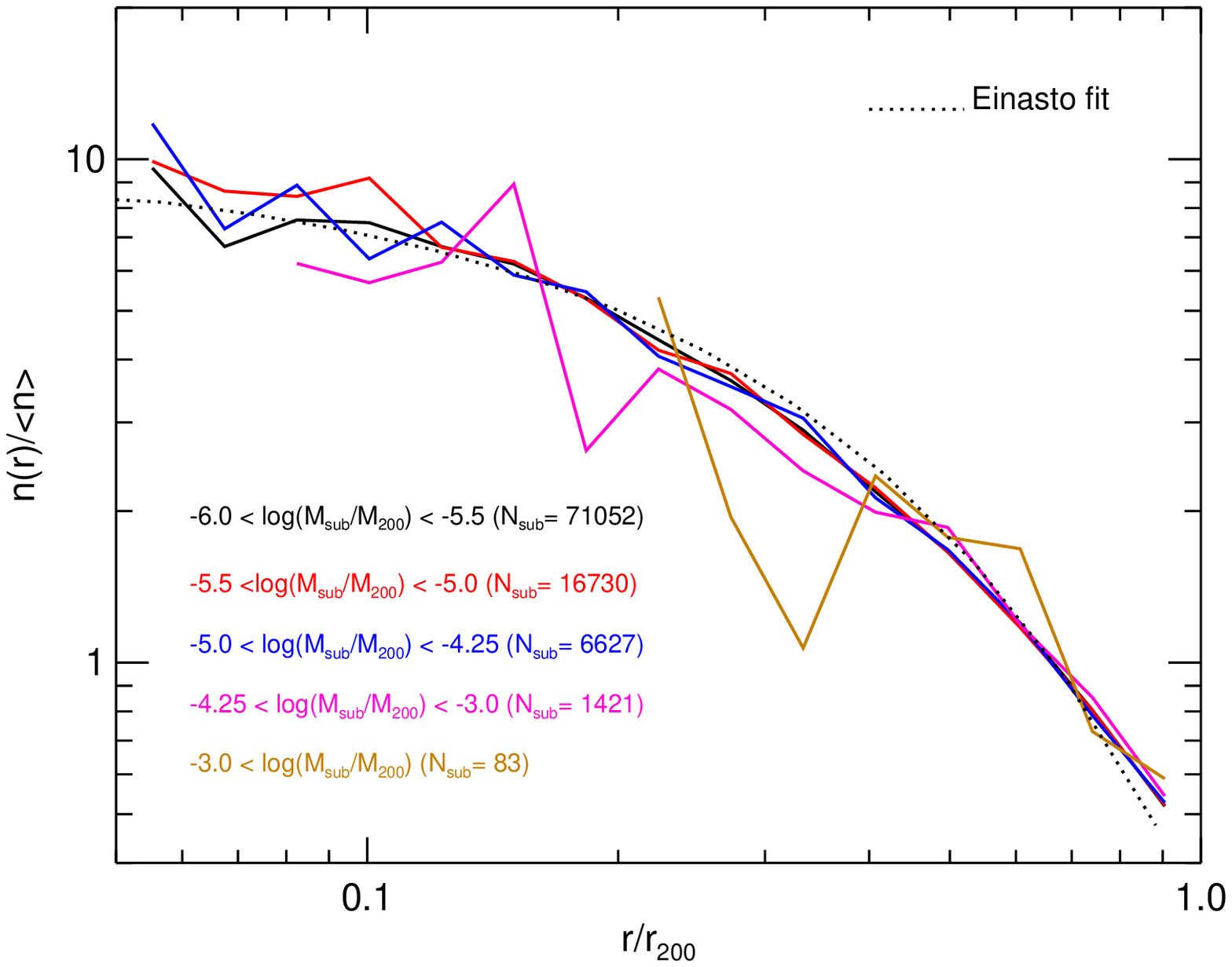}}
\caption {Subhalo number density profiles. The panel on the left shows
  the spatial distribution of subhaloes with more than $10$0 particles
  in each of the $9$ Phoenix level-2 clusters. Each profile is
  normalized to the mean number density of subhaloes within the virial
  radius. The thick dashed black curve traces the result of stacking
  all 9 level-2 Phoenix haloes. The profile obtained after stacking all
  level-2 Aquarius haloes is shown by the red dashed curve. Note that
  subhaloes are slightly more concentrated in the case of Phoenix than
  of Aquarius. The panel on the right shows the density profile of
  subhaloes in different bins of subhalo mass, computed after stacking
  all 9 level-2 Phoenix clusters. Note that the spatial distribution
  of subhaloes is approximately independent of subhalo
  mass.}
\label{FigSubhDensProf}
\end{figure*}

\subsection{Spatial Distribution}
\label{SecSubsDist}

The distribution of subhaloes within the main halo has been the subject
of many studies
\citep[e.g.][]{Ghigna2000,Diemand04a,DeLucia2004,Gao04a,Gao04b,sp08b,Ludlow2009}
over the past decade. This work has demonstrated that substructure
does not follow the same spatial distribution as the dark matter:
subhaloes tend to populate preferentially the outskirts of the main
halo and their spatial distribution is much more extended than the
mass. It also hinted that the number density profile of subhaloes is
roughly independent of subhalo mass, at least in the subhalo mass
range where simulations resolve them well and where they exist in
sufficient numbers for their spatial
distribution to be determined. This result has been confirmed recently by the Aquarius
simulation suite for haloes similar to the Milky Way \citep{sp08b}.

A number of observational diagnostics depend on the spatial
distribution of substructure, and it is therefore important to verify
that this result holds also on galaxy cluster scales. For example, recent
analyses indicate that total flux of dark matter annihilation
radiation is expected to be dominated by low-mass subhaloes
\citep{Kuhlen2008,sp08a,Gao2011}. It is therefore crucial to constrain
their spatial distribution in order to understand the expected angular
distribution of the annihilation flux and to design optimal filters to
aid its discovery \citep[see, e.g.,][]{Pinzke2011,Gao2011}.

We show the number density profile of subhaloes in
Fig.~\ref{FigSubhDensProf}. The left panel shows the profiles for each
of the 9 level-2 Phoenix haloes (thin lines), as well as the profile
corresponding to stacking all 9 haloes after scaling them to the virial
mass and radius of each cluster (thick dashed black curve). All
subhaloes with more than $100$ particles have been used for this plot.
This figure clearly confirms the results of earlier work: the subhalo
distribution is more extended than that of the dark matter; In
addition there is a well defined ``core'' in the central density of
the subhalo distribution; Subhaloes primarily populate the outskirts
of the main halo.

There is also considerable halo-to-halo scatter, especially near the
centre, where the number density of subhaloes may vary by up to a
factor of three. Comparing the average number density profile of
Phoenix with that of Aquarius (thick red dashed curve) reveals that
cluster subhaloes are slightly more abundant near the centre, by up to
$50\%$ at $r=0.1\, r_{200}$. In the outskirts of the main halo both
Aquarius and Phoenix give similar results. As discussed by
\citet{Ludlow2009}, the number density profile can be fitted
accurately by an Einasto profile (eq.~\ref{EqEinasto}), just like the
dark matter, but with quite different shape parameters: $\alpha\sim 1$
for subhaloes but $\sim 0.2$ for the main halo. An Einasto fit
to the Phoenix subhalo profile yields $r_{-2}=0.58  \, r_{200}$ and
$\alpha=1.0$. For Aquarius, the same procedure yields $r_{-2}=0.64\,
r_{200}$ and $\alpha=1.0$, and a central density normalization lower
by a factor of $1.3$, when expressed in units of $\langle n \rangle$, the
mean number density of subhaloes within $r_{200}$.

\begin{figure}
\hspace{0.13cm}\resizebox{8cm}{!}{\includegraphics{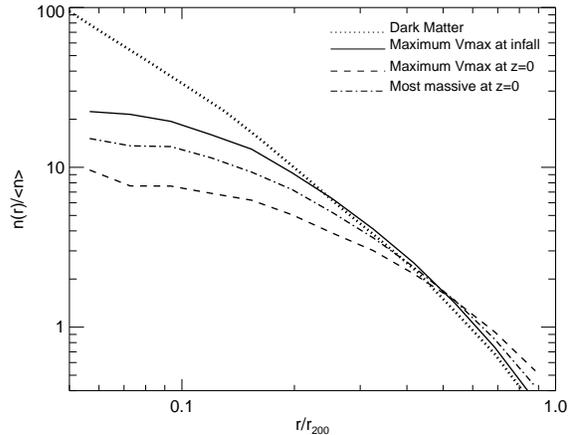}}
\caption{ Stacked subhalo number density profiles as a function of $r/r_{200}$
  for the nine Phoenix haloes and for different definitions of the lower
  subhalo ``mass'' limit. The solid line shows the radial profile for all
  subhaloes whose progenitors had a maximum circular $V_{\rm max}$ exceeding
  $45\, {\rm km~ s^{-1}}$ when they first fell into the cluster; the dot-dashed
  line shows a similar profile but for subhaloes with $V_{\rm max}$ greater
  than $30\, {\rm km~s^{-1}}$ at the present day; finally, the dashed line show
  the profile for all subhaloes containing more than 200 bound particles. For
  comparison, a dotted line shows the stacked dark matter mass profile of the
  clusters. The profiles are normalised to integrate to the same value within
  $r_{200}$. Note that none of the subhalo profiles matches the shape of the
  dark matter profile within $0.25r_{200}$.}
\label{FigSubSelect}
\end{figure}

Simplified schemes for populating dark matter simulations with
galaxies make a variety of assumptions about how to assign galaxies to
subhaloes.  A number of authors have argued that although present
subhalo mass and maximum circular velocity are strongly affected by
tidal stripping and so are poor indicators of galaxy properties, the
mass or circular velocity at infall are plausibly much better
and give meaningful results when used in subhalo abundance matching analyses
\citep{Vale2004, Conroy2006,Behroozi2010,Guo2010}. We study this issue
in Fig.~\ref{FigSubSelect}, which shows stacked number density
profiles for subhalo samples defined above thresholds in present mass,
present circular velocity and infall circular velocity. Note that
these thresholds are chosen so that each sample contains roughly
$6000$ subhaloes.  In agreement with earlier work, we see that sample
definition has a substantial effect on the inferred radial profile of
the subhalo population. Subhalo samples defined by present mass have
shallower profiles than samples defined by present circular velocity 
which, in turn, have shallower profiles than samples defined by infall
circular velocity. Note, however, that all these profiles differ
substantially from the mean dark matter density profile, especially in
the inner regions ($r<0.25\, r_{200}$), whereas observations show the
mean galaxy number density profiles in the inner regions of clusters
to follow the mean dark matter profiles quite closely
\citep[e.g.][]{Carlberg1997, Biviano2003, Sheldon2009}. Semi-analytic
models which explicitly follow the formation of galaxies within the
evolving subhalo population provide a better match to the observed
inner profiles because they include a population of ``orphan''
galaxies whose dark matter subhaloes have already been tidally
destroyed \citep{Gao04b,Wang2006,Guo2011}.

\begin{figure*}
\hspace{0.13cm}\resizebox{8cm}{!}{\includegraphics{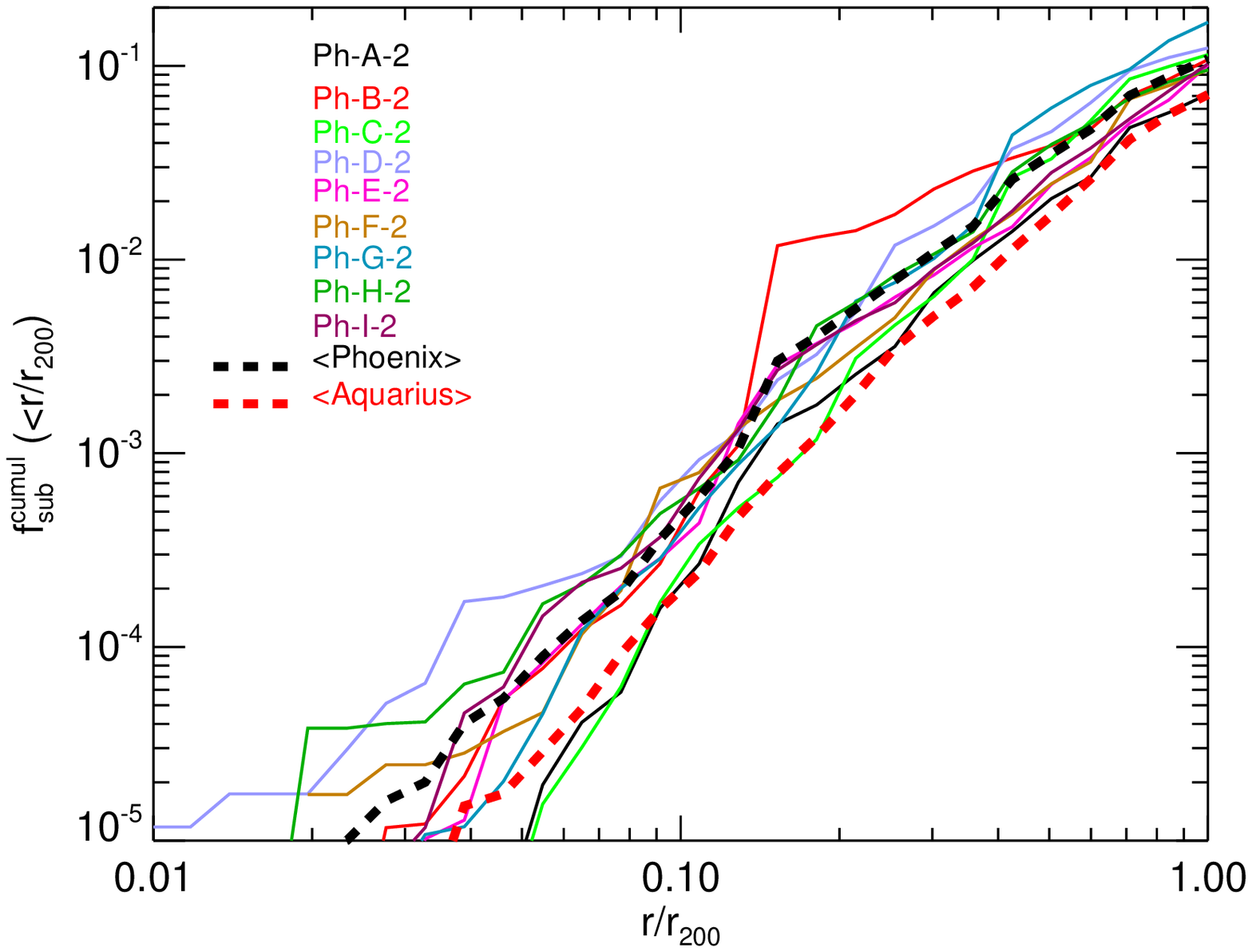}}
\hspace{0.13cm}\resizebox{8cm}{!}{\includegraphics{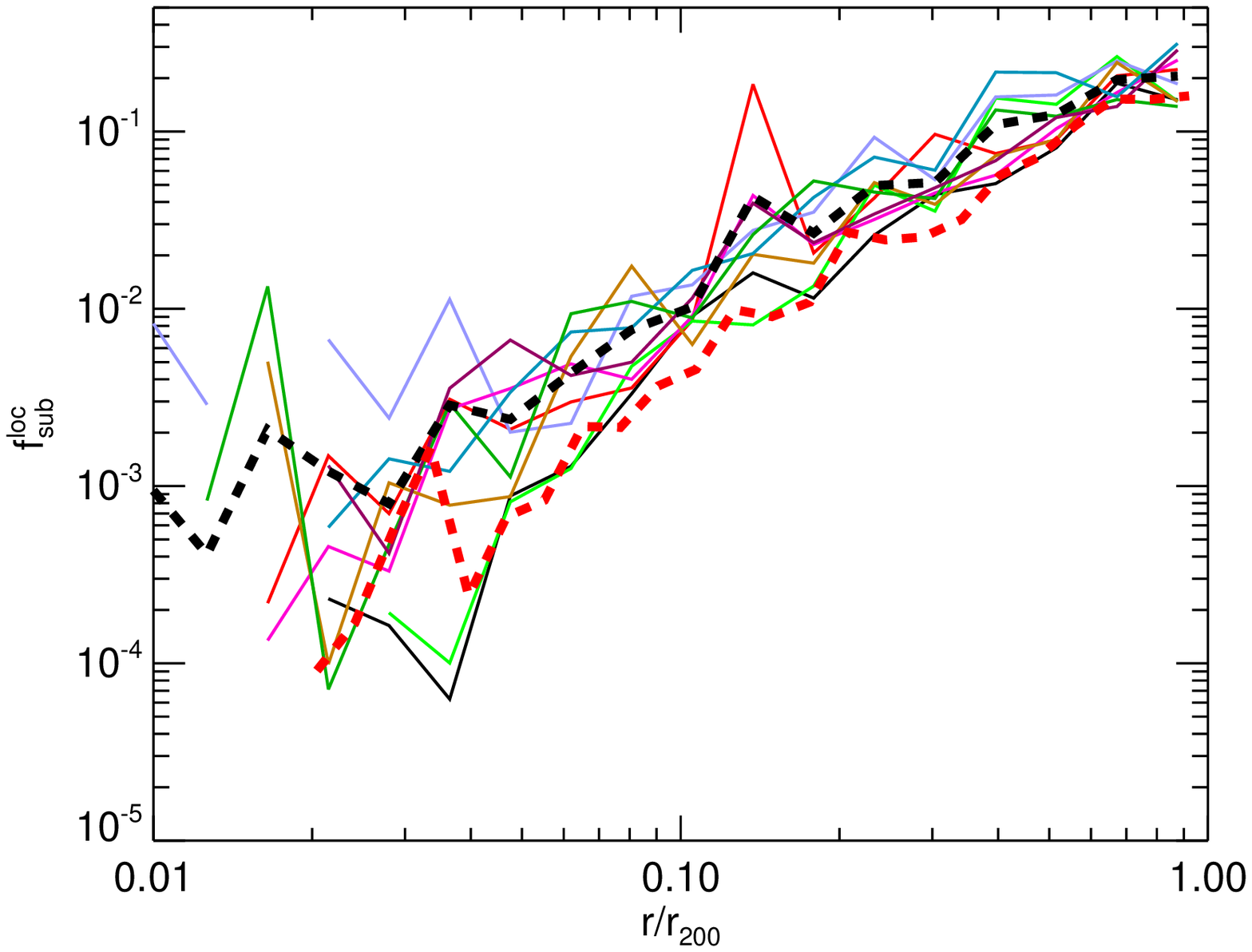}}
\caption{{\it Left panel}: Cumulative fractional contribution of
  subhaloes (resolved with more than $100$ particles) to the enclosed
  mass, shown as a function of radius for all level-2 Phoenix clusters
  (thin lines). A thick dashed black curve shows the average trend,
  computed after stacking all 9 Phoenix haloes. The corresponding
  result for Aquarius is shown by the thick dashed red curve. {\it
    Right panel}: Fraction of total mass contributed by substructure
  in different radial bins. As in the left panel, only subhaloes with
  more than $100$ particles are considered; black and red thick dashed
  lines correspond to the average trend computed after stacking all
  level-2 Phoenix and Aquarius haloes, respectively.}
\label{FigSubsCont}
\end{figure*}

Fig.~\ref{FigSubsCont} shows the fractional contribution of
substructure to the total mass of the halo, as a function of radius,
either in cumulative (left panel) or differential (right
panel) form. This figure shows quantitatively that substructure
contributes only a small fraction of the halo mass. This contribution
peaks in the outer regions; it is only $0.1\%$ at $r=0.02\, r_{200}$
but it reaches $10$-$20\%$ at the virial radius. The total mass
contribution is, on average, just over $10\%$ (see also
Table~\ref{TabHaloParam}). Results for Phoenix are similar to
Aquarius, adjusted up by a modest amount that reflects the overall
larger substructure fraction present in clusters relative to
galaxy-sized haloes. This adjustment is mainly noticeable in the inner
regions, reflecting our earlier conclusion that substructure in
Phoenix is more centrally concentrated than in Aquarius.

\section{Summary and Conclusions}
\label{SecConc}

We present the Phoenix Project, a series of simulations of the
formation of rich galaxy cluster halos in the $\Lambda$CDM
cosmogony. Phoenix simulations follow the dark matter component of $9$
different clusters with numerical resolution comparable to that
of the Milky Way-sized haloes targeted in the Aquarius Project
\citep{sp08b,Navarro10}. We report here on the basic structural
properties of the simulated clusters and their substructure, and
compare them with those of Aquarius haloes in order to highlight the
near mass invariance of cold dark matter haloes in the absence of
baryonic effects. Our main results may be summarized as follows.

{\tt Radial Profiles.} The recent formation of galaxy clusters causes 
many of them to be rapidly evolving and unrelaxed. This results in
mass profiles that are less well approximated by simple fitting
formulae such as the NFW or Einasto profiles than those of galaxy
haloes. Stacking clusters helps to average out inhomogeneities in the
mass distribution characteristic of transient states. The mass profile
of the stacked cluster is very similar to that of Aquarius haloes; it
can be well approximated by an Einasto profile, albeit with a slightly
larger value of the shape parameter, $\alpha$, and significantly lower
concentration. The similarity extends to the dynamical properties of
the haloes: when properly scaled, the average pseudo-phase-space
density and velocity anisotropy profiles of Aquarius and Phoenix
haloes are indistinguishable.

{\tt Density Cusp.} The central density cusp has, at the innermost
resolved radius ($r_{\rm conv}\sim 2\times 10^{-3} \,r_{200}$), an
average logarithmic slope $\langle \gamma \rangle=1.05 \pm 0.19$,
where the ``error'' refers to the halo-to-halo rms dispersion of the
$9$ level-2 Phoenix runs. This is only slightly steeper than that of
Aquarius haloes at comparable radii, for which $\langle \gamma
\rangle=1.01 \pm 0.10$). Although in some clusters $\gamma$ remains
roughly constant over a sizeable radial range near the centre, in the
majority of cases the profile keeps getting shallower all the way to
the innermost converged radius, with little evidence of convergence to
an asymptotic power-law behaviour.

{\tt Projected Profiles.} Because of their aspherical nature, the
surface density of Phoenix haloes varies greatly depending on the line of
sight, in some cases by up to a factor of $\sim 3$ at a given projected
radius. This affects especially the inner regions and may give rise to
substantially biased estimates of a cluster's total mass and
concentration. For example, NFW fits to the inner $500 \, h^{-1}$ kpc of
9 Phoenix haloes, on average, lead to estimates of $M_{200}$ and $c$
that can be overestimated by $20\%$ and $80\%$, respectively, when the
cluster is projected along the major axis and underestimated by $30\%$
and $20\%$ respectively when seen along the minor axis. The {\it shape} of
the surface density profile, on the other hand, is hardly affected by
projection. The average logarithmic slope of the surface density
profile declines gradually towards the centre, from $\langle \gamma_p
\rangle=0.35\pm 0.091$ at $R=10 \, h^{-1}$ kpc to $0.21\pm 0.054$ at
$R=3 \, h^{-1}$ kpc, again with no clear sign of approaching a power-law
asymptotic behaviour.

{\tt Substructure Mass Function.} Substructure is more abundant (by
about $\sim 35\%$ on average) in Phoenix clusters than in galaxy
haloes. At a given $\mu=M_{\rm sub}/M_{200}$, the cumulative number of cluster
subhaloes is higher in Phoenix by about $\sim 30\%$ compared to
Aquarius, with a tendency for the 
excess to increase at the low-mass end. In some cases the subhalo mass
function is best approximated by a power law with the critical slope
$N\propto \mu^{-1}$. There is significant halo-to-halo scatter,
however, and the average trend is subcritical. In the range $1\times
10^{-6}<\mu<1\times 10^{-4}$ we find that $N=0.010  \,\mu^{-0.98}$ fits well
the composite subhalo mass function of the $9$ level-2 Phoenix
clusters stacked together. For comparison, the same procedure for the
Aquarius haloes yields a very similar result: $N=0.012 \, \mu^{-0.94}$.

{\tt Substructure Spatial Distribution.}  We confirm earlier reports
that subhaloes are biased tracers of the halo mass distribution,
avoiding the central regions and increasing in prevalence gradually
from the centre outwards. As in galaxy haloes, the subhalo number
density profile appears to be independent of subhalo mass, and may be
approximated accurately by an Einasto profile, but with scale radius
$\sim 0.58\, r_{200}$ and a shape parameter much greater than that
of the dark matter distribution, $\alpha\sim 1.0$. Phoenix subhaloes
are slightly more concentrated than those of Aquarius
haloes: inside $0.1 \, r_{200}$ they make up roughly $0.05\%$ of the
enclosed mass, a factor of $2$ to $3$ times larger than in Aquarius
haloes. The difference decreases with increasing radius; in total
Phoenix subhaloes make up on average $11\%$ of the total mass, compared
with $7\%$ for Aquarius.

Our analysis confirms the remarkable structural similarity of CDM
haloes of different mass, whilst at the same time emphasizing the
small but systematic differences that arise as halo mass increases
from galaxies to clusters. Many of these differences may be ascribed
to the dynamical youth of galaxy clusters, which lead to larger
deviations of individual clusters from the average
trends. This argues for combining the results of as many clusters as
possible in order to average over the transient features of individual
systems and to uncover robust trends that may be fruitfully compared
with the predictions of the $\Lambda$CDM paradigm.

\section*{Acknowledgements}
Phoenix is a project of the Virgo Consortium. Most simulations were
carried out on the Lenova Deepcomp7000 supercomputer of the super
Computing Center of Chinese Academy of Sciences, Beijing, China, and
on Cosmology machine at the Institute for Computational Cosmology
(ICC) at Durham. The Cosmology machine is part of the {\small DiRAC}
facility jointly founded by {\small STFC}, the large facilities
capital fund of {\small BIS}, and Durham University. LG acknowledges
support from the one-hundred-talents program of the Chinese academy of
science (CAS),  the National Basic Research Program of China (program
973 under grant No. 2009CB24901), {\small NSFC} grants (Nos. 10973018
and 11133003), {\small MPG} partner Group family, and an {\small STFC}
Advanced Fellowship,  as well as the hospitality of the Institute for
Computational Cosmology at Durham University. CSF acknowledges a Royal Society
Wolfson Research Merit Award and ERC Advanced Investigator grant,
COSMIWAY. This work was supported in part by an 
{\small STFC} rolling grant to the ICC. We thank the anonymous referee
for a thoughtful and useful report.
\label{lastpage}


\bibliographystyle{mnras}
\bibliography{paper}

\section{Appendix}
\label{SecApp}

\subsection{Fitting formulae}

The fitting formulae used to describe the mass profile of Phoenix
haloes are the following: (i) The NFW profile \citep{NFW96,NFW97}, given by
\begin{equation}
\rho(r)=\frac{\rho_s}{(r/r_s)(1+r/r_s)^2},
\label{EqNFW}
\end{equation}
and (ii) the Einasto profile \citep{Einasto65},
\begin{equation}
\ln (\rho(r)/\rho_{-2})=(-2/\alpha)[(r/r_{-2})^{\alpha}-1].
\label{EqEinasto}
\end{equation}

Because these formulae define the characteristic parameters
in a slightly different way, we choose to reparametrise them in terms
of $r_{-2}$ and $\rho_{-2}\equiv \rho(r_{-2})$, which identify the
``peak'' of the $r^2\rho$ profile shown in the left panel of
Fig.~\ref{FigConvDensProf}.  This marks the radius where the logarithmic
slope of the profile, $\gamma(r)=-d\ln \rho/d \ln r$, equals the
isothermal value, $\gamma=2$. We note that, unlike NFW, when $\alpha$
is allowed to vary freely the Einasto profile is a 3-parameter fitting
formula. 

\subsection{Fitting procedure}

We compute the density profiles of each halo in $32$ radial bins
equally spaced in $\log_{10} r$, in the range $r_{\rm conv}< r <r_{\rm 200}$. All
haloes are centred at the minimum of the gravitational potential.
Best-fit parameters are found by minimizing the deviation between
model and simulation across all bins in a specified radial range.  In
the case of the density profile, the best fit is found by minimizing
the figure-of-merit function, $Q^2$, defined by
\begin{equation} 
Q^2={1 \over N_{\rm bins}} \sum_{i=1}^{N_{\rm bins}}
(\ln{\rho_i} - \ln{\rho_i^{\rm model}})^2.  \label{eq:q2}
\end{equation} 

This function provides a simple measure of the level of disagreement
between simulated and model profiles. It is dimensionless; it weights
different radii logarithmically; and, for a given radial range, $Q^2$ is
roughly independent of the number of bins used in the profile. The
actual value of $Q$ is thus a reliable and objective measure of the
average per-bin deviation from a particular model. Thus, minimizing
$Q^2$ yields for each halo well-defined estimates of a model's
best-fit parameters. The values of $Q_{\rm min}$ for each halo are
given in Table~\ref{TabHaloParam} for both Einasto and NFW fits.

It is less clear how to define a goodness-of-fit measure associated
with $Q^2$ and, consequently, how to assign statistically-meaningful
confidence intervals to the best-fit parameter values.  We have
explored this issue in \citet{Navarro10} and we refer the interested
reader to that paper for details. 

\end{document}